\documentclass[journal=jacsat,manuscript=article]{achemso}

\usepackage{amsmath,amsfonts,amssymb}
\usepackage{graphicx}
\usepackage{setspace}
\usepackage{tocloft}
\usepackage{orcidlink}
\usepackage{booktabs}
\usepackage{lscape}
\usepackage[]{braket}
\usepackage{simplewick}
\usepackage{longtable}
\usepackage{rotating}
\usepackage{pdflscape}  
\usepackage{geometry}
\usepackage{makecell}
\usepackage{xcolor}

\usepackage[version=3]{mhchem} 

\author{Shamik Chanda}
\affiliation[IISER Kolkata]
{Department of Chemical Sciences, Indian Institute of Science Education and Research Kolkata, Mohanpur, Nadia-741246, West Bengal, India}
\author{Pratyush Bhattacharjya}
\affiliation[IISER Kolkata]
{Department of Chemical Sciences, Indian Institute of Science Education and Research Kolkata, Mohanpur, Nadia-741246, West Bengal, India}
\author{Avijit Sen}
\affiliation[Syamaprasad College]
{Department of Physics, Syamaprasad College, Kolkata-700026, West Bengal, India}
\author{Sangita Sen}
\email{sangita.sen@iiserkol.ac.in}
\affiliation[IISER Kolkata]
{Department of Chemical Sciences, Indian Institute of Science Education and Research Kolkata, Mohanpur, Nadia-741246, West Bengal, India}
\title[An \textsf{achemso} demo]
  {UGA-SSMRPT2 - A Multireference Perturbation Theory Predicting Accurate Electronic Excitation Energies in Diverse Molecular Systems}

\abbreviations{IR,NMR,UV}

\begin{document}


\begin{abstract}

UGA-SSMRPT2, the spin-free perturbative analogue of Mukerjee's State-Specific Multireference Coupled Cluster Theory (MkMRCC) is known to be successful for size-extensive and intruder-free construction of dissociation curves. This work demonstrates that UGA-SSMRPT2 is also an accurate and computationally inexpensive framework for computing excitation energies. The method achieves near-chemical accuracy for the vast majority of $\pi \to \pi^*$, $n \to \pi^*$, charge-transfer, valence-Rydberg and Rydberg excited states commonly used for benchmarking electronic structure theories for excited states. Our results demonstrate that UGA-SSMRPT2 excitation energies lie within 0.20 eV of EOM-CCSD and/or well-established theoretical best estimates often surpassing the popular MRPT2 approaches like NEVPT2, CASPT2, and MCQDPT while typically requiring smaller active spaces. Its state-specific formulation circumvents the well-known intruder-state problem and eliminates the need for empirical parameters such as IPEA shifts in CASPT2. This work proposes UGA-SSMRPT2 as a robust, and scalable approach for modeling challenging electronic excited states.
\end{abstract}
\section{Introduction}
Accurate prediction of low-lying electronic excitations remains a cornerstone challenge in theoretical chemistry. Excited states govern the optical and photophysical behavior of molecules in photovoltaics\cite{khan20,taouali18}, light-emitting diodes\cite{forrest04,mitschke00}, photocatalysis\cite{Dabestani1998}, and photobiological processes\cite{li16,kundu09} and play important roles in technology. 
For example, charge-transfer (CT)\cite{mulliken62} states are deliberately harnessed in organic photovoltaics\cite{yu95}, nonlinear optical materials\cite{bredas94} (e.g., push–pull dyes), organic field-effect transistors\cite{tseng12} (through band alignment), and Organic Light Emitting Diode (OLED) emitters. However, accurate descriptions of these CT excitations requires an accurate treatment of electron correlation together with the inclusion of long-range exchange. Moreover, organic chromophore units are typically medium-sized involving conjugation and heteroatoms, both of which typically increase computational cost and often lead to multi-reference character. Rydberg states, which involve excitation to highly diffuse orbitals, are pivotal in spectroscopy and photoionization dynamics, often acting as intermediates that connect bound excitations with the ionization continuum. 
These states often involve intricate mixtures of single, double, Rydberg and charge-transfer excitations, requiring electronic structure methods that can capture both static and dynamic correlation with balanced accuracy and feasible computational cost. Benchmarking has played a critical role in assessing and calibrating these methods. While experimental spectra are abundant, they typically yield broad solution-phase bands and 0–0 transition energies rather than vertical excitations, making direct comparison difficult without vibronic and solvent modeling. Vertical excitation energies can be extracted experimentally only in a few cases, usually small diatomics. Hence, theoretical benchmarks have become the standard, providing consistent molecular geometries and basis sets. 

Single-reference approaches such as the Equation-of-Motion Coupled Cluster family (EOM-CCSD, EOM-CCSDT)\cite{stanton1993,kucharski2001,kowalski2001}, the related but partly perturbative family (CC2,CC3)\cite{christiansen1995,christiansen1995a,koch1997}, and the Algebraic Diagrammatic Construction family (ADC(2), ADC(3), ADC(4))\cite{harbach2014,dreuw2015} are recognized for their benchmark accuracy in single-excitation dominated systems. However, either their steep computational scaling (N$^6$–N$^7$) or the limitations of the single reference framework restricts their use for extended $\pi$-conjugated or strongly correlated systems, which constitute the vast majority of organic chromophores. Time-Dependent Density Functional Theory (TD-DFT)\cite{casida1995,gross1984,marques2004,burke2005,casida2012,ullrich2012}, though remarkably efficient and typically achieving mean absolute errors of $\sim$0.30 eV for valence states\cite{jacquemin2009,leang2012}, struggles with long-range charge-transfer\cite{dreuw2003,tawada2004,maitra2022_double_ct_tddft} (CT) and Rydberg excitations\cite{li2014,zhao2006} due to local exchange–correlation kernel limitations and the adiabatic approximation\cite{elliott2011,burke2005,maitra2016}. Even modern range-separated hybrids\cite{savin1996_recent_dft,refaelyabramson2015_prb_gwct} and double-hybrid functionals\cite{grimme2007}, while improving asymptotic behavior, introduce system-dependent parameters and remain inconsistent for multireference cases. Meanwhile, ADC(2) and CC2 methods achieve more consistent errors in excitation energies compared to TD-DFT. With computational scaling of $\mathcal{O}(N^5)$, they can be routinely applied to systems of up to $\sim$100 atoms and typically yield valence excitation energies within $\sim$0.10–-0.20 eV of experimental or high-level theoretical benchmarks\cite{hattig2000,winter2013,jacquemin2015,loos2018,loos2020}. Within the DFT framework, $\Delta$SCF with the maximum overlap method (MOM)\cite{gilbert2008} is able to handle difficult electronic excitations via explicit orbital relaxation\cite{barca2018,chanda2025}, at the cost of spin contamination of open shell states. A new alternative is the Multiconfiguration Pair-Density Functional Theory (MC-PDFT)\cite{manni2014}, where preliminary benchmarks\cite{hoyer2016,ghosh2015,hoyer2015} have shown that MC-PDFT can reproduce excitation energies for valence, Rydberg, and charge-transfer states with accuracy comparable to Complete Active Space Perturbation Theory (CASPT2)\cite{andersson1990}, but at a significantly lower computational cost. Functional dependence continues to be a challenge for all DFT-based methods.

Multireference (MR) or multiconfigurational wavefunction theories\cite{Szabo2012modern,hegarty1979,lischka2018}, which capture both static and dynamic electron correlation, are widely used for describing excited states and providing high-quality reference data. These methods are built upon a Multiconfiguration Self-Consistent Field (MCSCF) framework—most commonly the Complete Active Space SCF (CASSCF)\cite{olsen1988,taylor1984,schmidt1998}—which accurately accounts for static correlation within the active space and treats singly and multiply excited configurations on equal footing, but neglects dynamic correlation. The latter is recovered through post-CASSCF treatments such as Multireference Configuration Interaction (MRCI)\cite{Werner1988AnIN}, Multireference Coupled-Cluster (MRCC)\cite{Jeziorski1981,Li1995UnitaryApproximations,Li1997ReducedStates,Mahapatra1999_mrccdev,MAHAPATRA1998AApplications,MAHAPATRA1998163, Evangelista2008,Maitra2012,Maitra2014,Sinha2012,Sinha2013,Sen2018,Sen2012,Shee2013,Sen2013}, or Multireference Perturbation Theory (MRPT)\cite{anderson1990,andersson1992,hirao1992,Kozlowski1994ConsiderationsTheory,Zaitsevskii1996a,Zaitsevskii1995,Hoffmann1996,angeli2001,angeli2001a,angeli2002,rolik2003,szabados2005,finley1998,shiozaki2011,battaglia2020,angeli2006,Granovsky2011ExtendedTheory,devarajan2008,Mahapatra1999_ptdev,mahapatra1999_ptapp,Mahapatra1999jpca,Mao2012,Mao2012b,Sen2015,Sen2015b}. These constitute the family of multireference correlation theories, which are called uncontracted if the ansatz for dynamic correlation excites out of the individual model functions and contracted if it excites directly out of the CASSCF states\cite{MAHAPATRA1998163}. While non-perturbative approaches like MRCI and MRCC offer benchmark accuracy, their computational cost limits their use to small and medium systems. In contrast, MRPT2, which combine a multiconfigurational reference with a second-order perturbative correction, provides a more affordable alternative—often cheaper than the underlying CASSCF. Although absolute state energies are not usually up-to-the-mark, energy differences that are primarily of spectroscopic and chemical interest are often of sufficient accuracy. 

The most influential MRPT2 formulations include CASPT2\cite{andersson1990,andersson1992}, N-Electron Valence Perturbation Theory (NEVPT2)\cite{angeli2001,angeli2001a,angeli2002}, Multi Configurational Quasi Degenerate Pertubation Theory (MCQDPT2)\cite{hirao1992}, and Mukherjee's State Specific Multi Reference Perturbation Theory (SSMRPT2/Mk-MRPT2)\cite{Mahapatra1999_ptdev,mahapatra1999_ptapp,Mao2012,Mao2012b,Sen2015,Sen2015b}. An alternative framework, Multiconfiguration Perturbation Theory (MCPT)\cite{rolik2003,szabados2005}, extends MRPT concepts to general multireference starting points. CASPT2\cite{anderson1990}, NEVPT2\cite{angeli2001} and MCQDPT2\cite{hirao1992} in their simplest versions are state-specific theories but multi-state extensions like 
Multi-State Complete Active Space second-order Perturbation Theory (MS-CASPT2)\cite{finley1998}, Extended Multi-State CASPT2 (XMS-CASPT2)\cite{shiozaki2011}, Extended Dynamically Weighted CASPT2 (XDW-CASPT2)\cite{battaglia2020}, Quasi-Degenerate N-Electron Valence State Perturbation Theory (QD-NEVPT2)\cite{angeli2006}, and Extended Multi-Configuration Quasi-Degenerate Perturbation Theory (XMCQDPT2)\cite{Granovsky2012} exist. However, only NEVPT2 is intruder free and size-consistent in both state-specific and multi-state versions (QD-NEVPT2)\cite{angeli2006}. NEVPT2 offers several variants — Partially Contracted (PC-NEVPT2), {\color{black} also called Fully Internally Contracted (FIC-NEVPT2)}, Strongly Contracted (SC-NEVPT2)\cite{angeli2001a}, \cite{angeli2002}, and Uncontracted (UC-NEVPT2). The less contracted versions often provide higher accuracy by including a larger set of perturbers. The X versions of CASPT2 and MCQDPT2 also use partial decontraction to achieve higher accuracy. CASPT2 which employs a Fock-operator-based partitioning of the Hamiltonian, faces three classic challenges\cite{andersson1994,andersson1995,roos1995}: (i) intruder-state divergences from small energy denominators, (ii) state mixing and root flipping near degeneracies, and (iii) imbalance between open- and closed-shell zeroth-order Hamiltonians. Practical fixes—level shifts\cite{roos1995,forsberg1997}, Ionization Potential Electron Affinity (IPEA)\cite{ghigo2004} modifications, and multistate (MS-, XMS-, or XDW-) formulations—improve stability but introduce parameters and compromise formal size-extensivity. Static-Dynamic-Static PT (SDS-PT2)\cite{liu2014} is designed to avoid arbitrary level shifts in a framework similar to CASPT2. MCQDPT2, based on the Moller-Plesset (MP)\cite{moller1934}/ Epstein–Nesbet (EN)\cite{epstein1926,nesbet1955} partitioning, improves state-specificity but remains vulnerable to intruders and lacks strict size extensivity. 
\textcolor{black}{Alternative dynamically weighted multi-state MRPT2 formalisms have also been developed within the driven similarity renormalization group (DSRG) framework\cite{chenyangli2019}, and state-averaged DSRG-MRPT2 methods have subsequently been benchmarked for vertical excitation energies\cite{loos2019double}.}
The strength of NEVPT2 lies in its use of the Dyall Hamiltonian\cite{dyall1995} (two-body $H_0$) rather than the effective one body $H_0$'s of CASPT2 and MCQDPT2, which guarantees positive denominators and makes the theory intruder-free and rigorously size-extensive. Although it tends to slightly overestimate excitation energies for charge-transfer states, its strongly and partially contracted variants (SC- and PC-NEVPT2) typically achieve mean absolute errors near 0.1 eV for valence excitations\cite{sarkar2022}. CASPT2 has become the most widely used multireference method for excited states, particularly in medium-sized chromophores\cite{serrano-andres1993,serrano-andres1993a,serrano-andres1993b,serrano-andres1995,roos1996,serrano-andres1996,serrano-andres1996a,roos1999,serrano-andres2005}, transition-metal complexes\cite{fouqueau2005,neese2006}, photobiological chromophores\cite{siegbahn2003,schapiro2010}, charge-transfer systems\cite{serrano-andres1998,ghosh2015}, and widely studied dyes such as BODIPYs\cite{wen2018} and iron-based coordination complexes\cite{pierloot2006,suaud2009,kepenekian2009}, beyond the capacity of EOM-CC, CCn or ADC(n) owing to remarkably advanced algorithms and implementations.

SSMRPT2\cite{Mahapatra1999_ptdev,mahapatra1999_ptapp,Mao2012,Mao2012b,Sen2015,Sen2015b} pioneered by Mukherjee and co-workers is a highly successful but less explored MRPT2. It is the perturbative analogue of Mukherjee's state-specific multireference coupled-cluster (SS-MRCC or Mk-MRCC)\cite{Mahapatra1999_mrccdev,MAHAPATRA1998AApplications,MAHAPATRA1998163,Evangelista2008,Maitra2012,Maitra2014,Sinha2012,Sinha2013} theory, which uses a one-body Fock like $H_0$ and retains the formal rigor size extensivity, intruder free, and explicit state specificity of the parent theory while offering MRPT-like efficiency. The strength of SSMRPT2 has been primarily demonstrated in the context of construction of Potential Energy Curves (PECs) owing to its intruder-free nature\cite{Mao2012,Mao2012b,Sen2015,Sen2015b}. We extend this exploration to the computation of electronic excitation energies in this work and show its efficacy in providing accurate excitation energies with much smaller active spaces than advocated by CASPT2 which is by far the most popular MRPT theory for excitation energies. 

Over the past two decades, extensive benchmark efforts have established reliable theoretical best estimates (TBEs)\cite{hattig2000,schreiber2008,silva-junior2010,silva-junior2010b,winter2013,jacquemin2015,loos2018,loos2020,veril2021,kannar2014,kannar2017,loos2021,loos2025}. Theories can be evaluated against benchmark values for a well-chosen and curated 'test set' of molecules in terms of statistical quantities like mean deviation (MD), mean absolute deviation (MAD), root mean-square deviation (RMSD), standard deviation ($\sigma$), maximum absolute deviation (Max $|\Delta|$), Pearson correlation coefficient (R) and coefficient of determination (R$^2$). The definitions are provided in the supporting information (SI). Among the benchmark databases, Thiel’s set\cite{schreiber2008,silva-junior2010b,silva-junior2010a} remains the most influential. In 2008, Thiel and co-workers reported $\sim$160 theoretical best estimates (TBEs) for 28 organic molecules ($\sim$90 excited states), primarily derived from CASPT2/TZVP for singlets and CC3/TZVP for triplets\cite{schreiber2008}. In 2010, they updated the dataset with aug-cc-pVTZ values, adopting CC3 as the standard for singlets also, thereby exposing the pronounced basis-set dependence of CASPT2\cite{silva-junior2010b}. González and collaborators\cite{zobel2017}, showed that CASPT2 without the IPEA shift yields the smallest errors at the double-$\zeta$ level, while triple- and quadruple-$\zeta$ basis sets benefit from applying IPEA = 0.25. However, for larger molecules and transition-metal complexes, the IPEA shift can lead to systematic overestimation of excitation energies, highlighting its empirical and system-dependent character. This nuance is central to practical CASPT2 applications, as the choice of IPEA directly influences quantitative reliability\cite{ghigo2004}. Subsequent benchmark studied by Loos and Jacquemin extended the chemical diversity of the test set and proposed TBEs using selected configuration interaction (CI)\cite{garniron2018,scemama2019} and high-level coupled-cluster (CC) methods—-selected CI\cite{garniron2018} for systems with up to three non-H atoms, CCSDTQ\cite{kucharski1991,kallay2004,hirata2004} for four, and CCSDT\cite{noga1987,scuseria1988,kucharski2001} for 5–10 non-H atoms. Their “mountaineering” benchmarks, combining selected CI with millions of determinants and near-full CI reference data (up to CCSDTQP), yielded nearly exact excitation energies of 110 singlet and triplet valence and Rydberg states for 18 small CNOH molecules\cite{loos2018}.
Later, a follow-up study\cite{loos2020} reported TBEs for additional 27 organic chromophores with 4--6 non-H atoms using CC3, CCSDT-Q or selected CI calculations. These results established CCSDTQ as virtually indistinguishable from FCI (MAD$\sim$0.01 eV), while iterative triples methods such as CC3 and CCSDT-3 achieved MADs near 0.03 eV. In contrast, ADC(3) tended to overcorrect ADC(2)\cite{loos2020b}, showing less systematic accuracy. Later studies reinforced these conclusions. Schwabe and Goerigk\cite{schwabe2017} replaced experimental reference values with CC3 data, creating benchmarks that serve as de facto standards for TDDFT validation. Watson et al. (2013) demonstrated near-identical excitation energies from CCSDT-3 and CC3\cite{watson2013}, while Nooijen and co-workers analyzed subtle differences between CC3 and CCSDT-3\cite{demel2013}. Dreuw and collaborators confirmed that ADC(3) does not consistently surpass CC3\cite{harbach2014}, and Kannar and Szalay provided additional CCSDT benchmarks with TZVP and aug-cc-pVTZ\cite{kannar2014,kannar2017}. In all cases, improvements beyond CC3 were marginal—for example, the $\pi \to \pi^*$ transition in ethylene appears at 8.37 eV (CC3/TZVP), 8.38 eV (CCSDT/TZVP), and 8.36 eV (CCSDTQ/TZVP), differing by less than 0.01 eV. Collectively, these results establish CC3 as the practical gold standard for excited-state benchmarking. Sarkar et al., recently benchmarked NEVPT2/aug-cc-pVTZ and CASPT2/aug-cc-pVTZ against the TBEs from Loos et al.\cite{loos2018,loos2020}, for 284 excited states across 35 small organic molecules (174 singlets, 110 triplets; 206 valence, 78 Rydberg; 78 $n \to \pi^*$, 119 $\pi \to \pi^*$, 9 double excitations), showing that FIC-CASPT2 with a standard IPEA shift and PC-NEVPT2 both deliver reliable vertical excitations with small, nearly uniform errors of $\sim$0.11 and 0.13 eV respectively\cite{sarkar2022}. This complements earlier benchmarks of single-reference methods and indicates that CASPT2, ADC(2), and EOM-CCSD achieve comparable reliability for states dominated by single excitations. Multireference states continue to offer significant challenges.

In this work, we benchmark Unitary Group Adapted State-Specific Multireference Perturbation Theory (UGA-SSMRPT2)\cite{Sen2015,Sen2015b} excitation energies against EOM-CCSD\cite{stanton1993} and high-level TBEs\cite{loos2018,loos2020,veril2021,loos2025} for a diverse set of systems—drawing from the test sets of Thiel et. al.\cite{schreiber2008,silva-junior2010b} and Loos et. al.\cite{loos2018,loos2020,sarkar2022} and a few additional charge-transfer(CT) systems--showing that it achieves near-chemical accuracy ($\sim$0.10--0.20 eV) with smaller active spaces than conventional MRPT2 variants. The remainder of this paper is organized as follows: Section II presents the theoretical framework and working equations of UGA-SSMRPT2; Section III describes computational details and benchmarking protocols with discussions of the results and comparative analyses; and Section IV concludes the study with a summary of the main findings and the possible theoretical improvements.

\section{UGA-SSMRPT2 Theory}
The CASSCF wave function for the $k$th state is written as\cite{olsen1988},
\begin{eqnarray}
  |\Psi_k ^0\rangle = \sum_{\mu=1}^{n_{\text{CAS}}} |\phi_\mu ^0
  \rangle c^0_{\mu k}, \label{CASfn}
\end{eqnarray}
where $|\Phi_\mu^0\rangle$ are configuration state functions (CSFs) and $c^0_{\mu k}$ their expansion coefficients. This expansion naturally spans single, double, and higher excitations within the active space, thereby treating static correlation rigorously and describing n-tuply excited states on an equal footing with singly excited ones. The relative magnitudes of the CI coefficients $c^0_{\mu k}$ serve as direct diagnostics of multireference character. However, CASSCF does not incorporate dynamic correlation, which must be added through post-CASSCF correlation treatments.

The UGA-SSMRPT2\cite{Sen2015,Sen2015b,chakravarti2021} theory originates from the Unitary Group Adapted State-Specific Multireference Coupled Cluster (UGA-SSMRCC or UGA-MkMRCC)\cite{Maitra2012,Sinha2012,Sinha2013,Maitra2014} formalism, which is the spin-free version of Mukherjee's SSMRCC (MkMRCC)\cite{Mahapatra1999_mrccdev,MAHAPATRA1998AApplications}. The family of UGA-MRCC theories also consists of multi-state variants such as the Unitary Group Adapted Open Shell Coupled Cluster (UGA-OSCC)\cite{Sen2018}, Unitary Group Adapted State Universal Multireference Coupled Cluster (UGA-SUMRCC)\cite{Sen2012,Shee2013,Sen2013}, Unitary Group Adapted Quasi Fock Multireference Coupled Cluster (UGA-QFMRCC)\cite{Shee2013,Sen2013}. \textcolor{black}{In this UGA framework, the CSF's, $\phi_\nu^0$, are specifically Gel'fand states\cite{Maitra2012} with their spin adaptation specified through the graphical unitary group approach (GUGA)\cite{Brooks1979}}.

In general, the exact correlated state can be expressed as the action of a state-specific wave operator on a CASSCF reference defined in Eq.~\ref{CASfn}:
\begin{eqnarray}
 |\Psi_k\rangle &= \Omega |\Psi^0_k\rangle &= \sum_\mu \Omega |\phi^0_\mu\rangle c^0_{\mu k}.
\end{eqnarray}
In UGA-MRCC theories, the operator $\Omega$ is defined in its uncontracted form as,
\begin{eqnarray}
  \Omega &= \sum_{\nu} \Omega_\nu &= \sum_\nu \{ e^{T_{\nu}} \} |\phi^0_\nu\rangle \langle\phi^0_\nu|,
\end{eqnarray}
with \{$\phi_\nu^0$\} being unitary-group-adapted configuration state functions 
of proper spin and spatial symmetry and \{$T_\nu$\} = $\displaystyle\sum_{l} \{T_\nu ^l\}$ being spin-free cluster operators that excite $\phi_\nu^0$ to all possible spin-free virtual functions \{$\chi_\nu^l$\}. \textcolor{black}{The cluster operators \{$T_\nu$\} are normal ordered with respect to the core (the common closed shell part of all $\phi_{\nu} ^0$s) and can thus have active orbital destruction operators if these orbitals are singly occupied in that $\phi_{\nu} ^0$.} The ansatz for UGA-SSMRCC is state-specific and therefore takes the form
\begin{eqnarray}
 |\Psi_k\rangle = \Omega_k |\Psi^0_k\rangle = \sum_\mu \{ e^{T_{\mu}} \} |\phi^0_\mu\rangle c^0_{\mu k},
\end{eqnarray}
for a specific k, \textcolor{black}{where the wave operator is an exponential of the \{$T_\mu$\} operators which is overall normal ordered again with respect to the core as vacuum allowing the cluster operators to commute among themselves.}

In the perturbative analogue, UGA-SSMRPT2, the $H_0$ is defined as a multi-partitioned Fock operator constructed as, 

\[
H_0^\mu = \sum_{p,q}\tilde{f}_{\mu p}^{\;q} =\sum_{p,q} [ f_{c\,p}^{\;q}
   + \sum_{u_d \in \mu}\bigl( 2\,V_{p u_d}^{\;q u_d} - V_{p u_d}^{\;u_d q}\bigr)
   + \sum_{u_s \in \mu} V_{p u_s}^{\;q u_s}],
\]

where \(f_c\) denotes the core Fock operator, which is common for all 
\(\phi_\mu s\). Indices \(p, q\) are general orbital indices but should belong 
to the same ``class,'' i.e., inactive holes (\(i,j,\ldots\)), 
inactive particles (\(a,b,\ldots\)), or active orbitals (\(u,v,\ldots\)) \textcolor{black}{with $u_d$ and $u_s$ referring to doubly and singly occupied active orbitals in a given model function $\phi_{\mu}$, respectively. $V$ refers to the two-body integrals labelled by spatial orbital indices.} 
The zeroth-order Hamiltonian $H_0$ is thus class-diagonal, and closely analogous to the Møller–Plesset\cite{moller1934} partitioning in the multireference setting. \textcolor{black}{In the SA-SSMRPT2 theories\cite{Mao2012,Mao2012b}, the $H_0$ is taken to be diagonal and the cluster operators for $\phi_\mu ^0$ are taken in normal order with respect to the largest closed shell part of that model function. Both Moller-Plesset (MP) and Epstein-Nesbet (EN) like partitioning have been explored in this context\cite{Mao2012b}. A comparison of the theoretical similarities and differences between UGA-SSMRPT2 and SA-SSMRPT2 has been presented by Sen et al.\cite{Sen2015}.}

Considering only first order terms from the Schr\"{o}dinger equation containing the ansatz in Eq.~(4), the generic form of the UGA-SSMRPT2 working equations can be written as:  
\begin{eqnarray}     
   \langle\chi_{\mu}^l|H|\phi_\mu^0\rangle c_\mu^0  
  + \langle\chi_{\mu}^l| 
    \contraction{}{H_0}{}{T^{(1)}_\mu} %
    \{ H_0 T^{(1)}_\mu \} |\phi_\mu^0\rangle c_\mu^0       
  - \sum_\nu \langle\chi_{\mu}^l| 
    \contraction{}{T^{(1)}_\mu}{}{W_{\nu\mu}} %
    \{ T^{(1)}_\mu W_{\nu\mu} \} |\phi_\mu^0\rangle c_\mu^0          
  + \sum_\nu \langle\chi_{\mu}^l| 
    \{ T^{(1)}_\nu - T^{(1)}_\mu \} |\phi_\mu^0\rangle H_{\mu\nu} c_\nu^0  
    = 0
   \label{pt_eqn}
\end{eqnarray}
where $|\chi^{l}_{\mu}\rangle$ is the spin-free excited–state CSF generated by the action of a spin-free excitation operator $\{E^{l}_{\mu}\}$ normal-ordered with respect to the common core function as $
|\chi^{l}_{\mu}\rangle = \{E^{l}_{\mu}\}\,  |\phi_{\mu}^0\rangle$. The cluster operators are thus, $T_{\mu} = \sum_{l} t^{l}_{\mu} \{E^{l}_{\mu}\}$ and $W_{\nu\mu}$ is any closed component of H$_0$ scattering between model functions $\mu$ and $\nu$. The derivation and other details can be found in the references\cite{Sen2015,Sen2015b,chakravarti2021}. The derivation ensures that the denominators of the cluster amplitudes remain intruder-free and the connected nature of the working equations ensures size extensivity.   
{\color{black}
Both these aspects are clearly described in Secs.~IIE and IIF by Maitra \textit{et al.}~\cite{Maitra2012} in the context of UGA-SSMRCC, and similar arguments hold for UGA-SSMRPT2, as discussed in Sen et al.\cite{Sen2015b}. The lack of invariance of the energy with rotation of the active orbitals is a significant disadvantage of the family of Hilbert space MRCC and MRPT theories stemming from the use of the Jeziorski-Monkhorst ansatz. UGA-SSMRPT2 has the same issue although the problem is less severe for excitation energies than for potential energy surfaces. We have used pseudo-canonical orbitals in our applications.

In particular, for UGA-SSMRPT2, Eq.~(\ref{pt_eqn}) may be rearranged to express the amplitude value as
\begin{equation}
t_{\mu}^{l}(\mu)
=
\frac{
H_{\mu}^{l}
+
\contraction{}{H_0}{}{T^{(1)}_\mu}
\{ H_0 T^{(1)}_\mu \}_\mu^l
-
\displaystyle\sum_{\nu \neq \mu}
\contraction{}{T^{(1)}_\mu}{}{W_{\nu\mu}}
\{ T^{(1)}_\mu W_{\nu\mu} \}_{\mu}^{l}
+
\displaystyle\sum_{\nu \neq \mu}
\left\{
T_{\nu}^{(1)}
\right\}_{\mu}^{l}
\, H_{\mu\nu}\,
\dfrac{c_{\nu}^{0}}{c_{\mu}^{0}}
}{
E_{\mathrm{CAS}}
-
H_{ll}^{0}
+
H_{\mu\mu}^{0}
-
H_{\mu\mu}
}
\end{equation}
where, the notation ($\mu$) in $t_\mu^l(\mu)$ represents the fact that $t_\mu^l(\mu)$ is the amplitude acquired by the operator $\{E_\mu^l\}$ when it acts on $|\phi_\mu^0\rangle$. Similarly, $\{T_{\nu}^{(1)}\}_{\mu}^{l} = \langle\chi_\mu^l|t_\nu^l(\mu)\{E_{\nu}^l\}|\phi_\mu^0\rangle$.

Since the difference
\(
H_{\mu\mu}^{0} - H_{\mu\mu}
\)
is expected to be small, and the CASSCF energy
\(
E_{\mathrm{CAS}}
\)
is well separated from
\(
H_{ll}^{\circ}
\),
the denominator remains intruder-free, provided the active space is judiciously chosen. However, the choice of active space becomes progressively more difficult for higher-lying excited states with the starting Hartree-Fock orbitals also becoming poorer guesses for making this choice. Thus, intruder states may re-emerge if very high-lying states are targeted.

Eq.~(\ref{pt_eqn}) can be solved for the set of variables
\(
\{ {t_{\mu}^{l}}^{(1)}\}
\).
The second-order effective Hamiltonian,
\(
\tilde{H}_{\mu\nu}^{[2]}
\),
can then be constructed as
\begin{equation}
\tilde{H}_{\mu\nu}^{[2]}
=
H_{\mu\nu}
+
\sum_{l}
H_{\mu l}
\, t_\nu ^{l^{(1)}} ,
\end{equation}
where the superscript $[2]$ denotes that all contributions up to second order are included.

The \textit{unrelaxed} energy is defined as the expectation value of
\(
\tilde{H}_{\mu\nu}^{[2]}
\)
with respect to the zeroth-order CASSCF wave function and is denoted throughout the paper as
\(
\langle \mathrm{UGA\text{-}SSMRPT2} \rangle
\),
\begin{equation}
\langle \mathrm{UGA\text{-}SSMRPT2} \rangle
= E^{[2]}_{unrelaxed} =
\sum_{\mu\nu}
c_{\mu}^{(0)}
\, \tilde{H}_{\mu\nu}^{[2]}
\, c_{\nu}^{(0)} .
\end{equation}

UGA-SSMRPT2 further allows for coupling between dynamical and static correlation effects by defining a so-called \textit{relaxed} energy, obtained through diagonalization of the effective Hamiltonian,
\begin{equation}
\sum_{\nu}
\tilde{H}_{\mu\nu}^{[2]}
\, c_{\nu}
=
E_{\mathrm{relaxed}}^{[2]}
\, c_{\mu} .
\end{equation}

The quantity
\(
E_{\mathrm{relaxed}}^{[2]}
\)
is referred to as the UGA-SSMRPT2 energy in the remainder of this work. The updated coefficients
\(
\{ c_{\mu} \}
\)
are subsequently used to recompute the amplitudes
\(
\{ t_{\mu}^{l} \}
\),
yielding relaxed amplitudes. A nested iterative procedure is carried out until self-consistency is achieved with respect to both the configuration-interaction coefficients and the perturbative amplitudes. UGA-SSMRPT2 and $\langle$UGA-SSMRPT2$\rangle$ have also been referred to as SSMRPT2 and $\langle$SSMRPT2$\rangle$ in shorthand. 
}

\textcolor{black}{On account of the decontracted nature of the cluster amplitudes in UGA-SSMRPT2, the number of sets of variables \{t$_\mu^l$\} scales as the number of model functions \{$\phi_\mu$\}, ie. $n_{CAS}$. The number of variables in each set also typically increases with increase in the number of active orbitals. The use of incomplete model spaces along with slight modification of the choice of cluster amplitudes is one way of reducing this scaling as discussed in~\cite{Sen2013,Mukhopadhyay1989Size-extensiveSpaces,meissner1989,meissner1990,Mukherjee1986,Mukherjee1986a,Mukherjee1988}. In our implementation, we only retain the model functions with CASSCF coefficients above a user-defined threshold which is a crude but effective way of eliminating unnecessary model functions. The computational cost of the amplitude equations themselves is low, scaling roughly as $N^5$ with the most expensive term (assuming $n_a > n_i > n_u$) having a scaling of $n_i^2n_a^3$ where $n_i$, $n_u$ and $n_a$ are the number of inactive core, active, and inactive virtual orbitals respectively. However, for a very large number of active orbitals and model functions, the construction of the two-body and three-body effective Hamiltonian components becomes significantly costly due to a contraction with a two-body and three-body reduced density matrix (RDM) with a scaling of $n_{CAS}^2n_u^4$ and $n_{CAS}^2n_u^6$ respectively. Here, $n_{CAS}$ itself scales as $n_u!$. Although the formal computational scaling of working equations in internally contracted state-specific theories is similar ($N^5$ or $N^6$ depending on the specific formulation) and the number of variables for each state increases with $n_u$, the number of sets of variables do not increase unlike that in UGA-SSMRPT2 providing some computational advantage. For FIC-CASPT2 the construction of 4-RDMs is often touted as the biggest computational bottleneck and forms the subject of many of the approximations and code optimizations\cite{phung2016,wouters2016}. Similar cost improvements may also be implemented for UGA-SSMRPT2.}

Compared to other MRPT2 approaches, UGA-SSMRPT2 possesses all the desirable features: avoid intruder states without empirical shifts (unlike CASPT2), preserve size-consistency and extensivity (unlike MCQDPT2, MRCI), and thus offers balanced treatment across valence, Rydberg, $n \to \pi^*$, $\pi \to \pi^*$ and charge-transfer states. Its unitary-group adaptation enforces rigorous spin adaptation, and the state-specific formulation prevents root-flipping issues in strongly multireference scenarios. 
\textcolor{black}{Furthermore, by construction, UGA-SSMRPT2 admits a systematic hierarchy of improvements with respect to the quality of $H_0$ and the reference description. While the method is systematically improvable with respect to the definition of $H_0$ or the perturbative order, accurate excitation energies are already achieved at second order with a multi-partitioned effective one-body Fock operator as $H_0$ and a minimal chemically relevant active space, as demonstrated in the present work. Systematic improvability with size of the active space refers to the ability to reduce errors arising from an insufficient model space while approaching the physically relevant minimal active space, rather than to the indiscriminate enlargement of the active space beyond the minimal chemically relevant manifold.}

\begin{figure}[h!]
    \centering
    \includegraphics[width=0.48\linewidth]{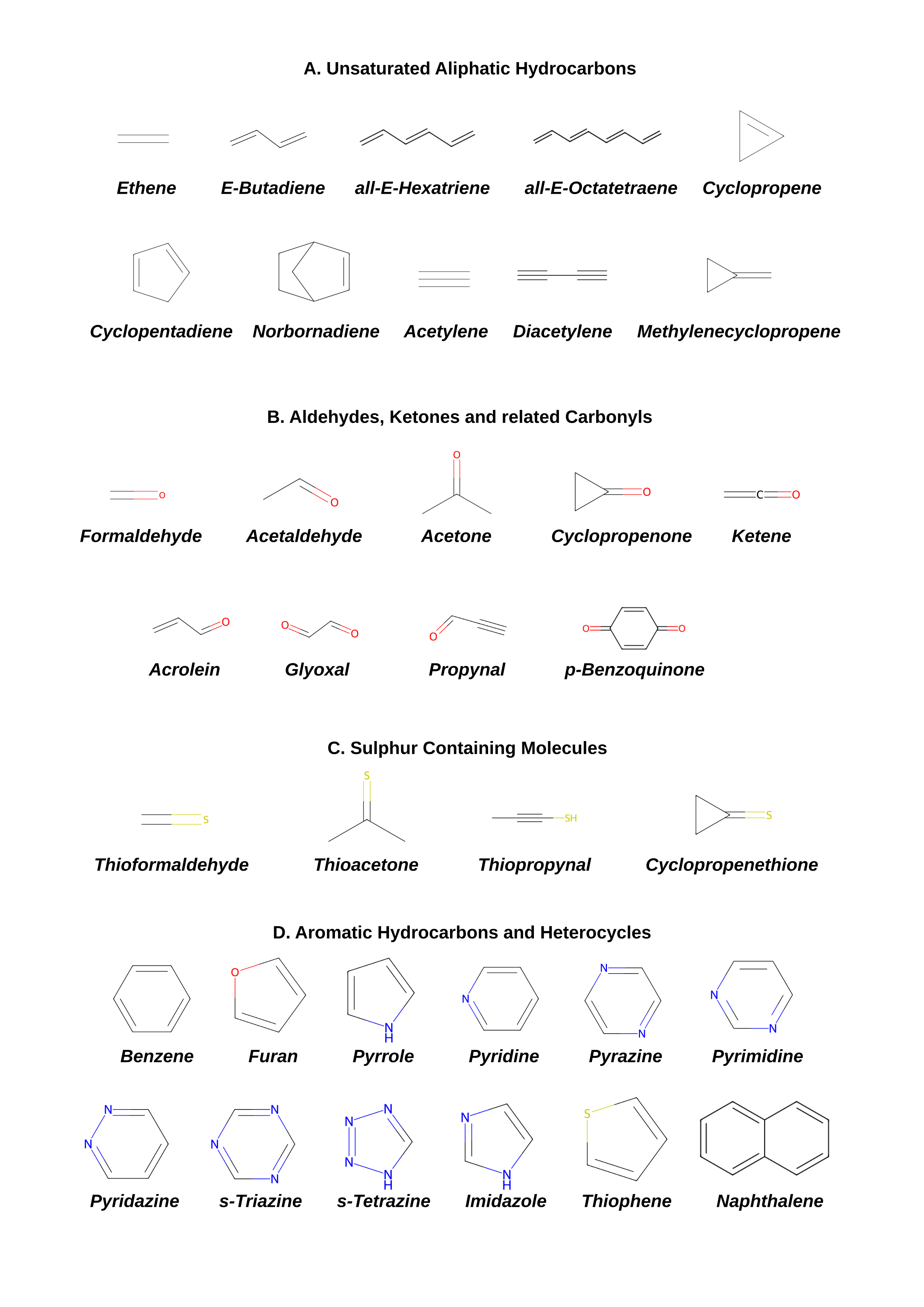}%
    \hfill
    \includegraphics[width=0.48\linewidth]{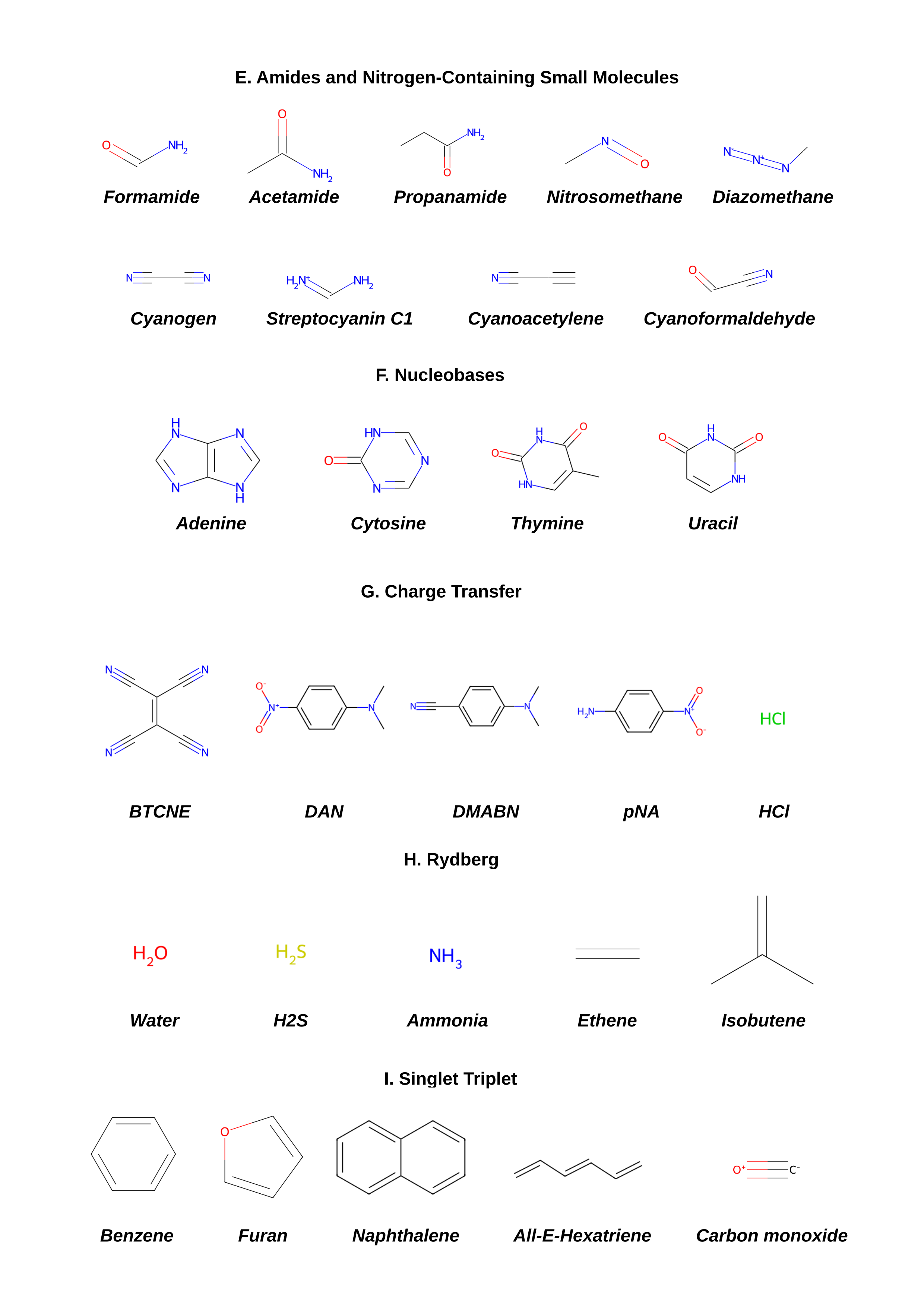}
    \caption{Test set of molecules in this study.}
    \label{fig:molecules}
\end{figure}

\section{Results and Discussions}
The UGA-SSMRPT2 code is an independent package interfaced with GAMESS-US\cite{Schmidt1993}. Molecular geometries for the Thiel set were taken from the supplementary material of Thiel \textit{et al.}\cite{schreiber2008}, whereas structures for the Loos set, Rydberg systems, CO, and HCl were obtained from the QUEST database\cite{loos2021}. Geometries of the charge-transfer (CT) complexes B–TCNE and DMABN were adopted from Refs.~\cite{stein2009,moore2013}, while those for pNA and DAN were optimized at the B3LYP/6-311G(d,p) level using Gaussian 09 following Ref.~\cite{sun2013}. The aug-cc-pVTZ basis set was used for Rydberg excitations in H\textsubscript{2}O, H\textsubscript{2}S, NH\textsubscript{3}, and ethene, and aug-cc-pVDZ for isobutene. CT excitation energies were computed with the cc-pVDZ basis set, except for HCl, where aug-cc-pVTZ was employed. All remaining excitations were evaluated using the def2-TZVP basis set, except for CO, for which aug-cc-pVTZ was used. Basis-set dependence for low-lying valence excitations is generally small ($<$0.1 eV)\cite{schreiber2008,silva-junior2010b}, although states exhibiting valence–Rydberg mixing may show larger variations; such cases are explicitly discussed where relevant. All calculations were performed in C\textsubscript{1} symmetry, i.e., without exploiting point-group symmetry, in order to access all states on equal footing. CASSCF, NEVPT2, and CASPT2 calculations were carried out using ORCA 6.0.1\cite{orca2020}, while MCQDPT2 computations used GAMESS-US. EOM-CCSD excitation energies were obtained using a trial version of Q-Chem 6.0.1\cite{epifanovsky2021_qchem5}.

The \textit{test set} comprises a chemically diverse collection of molecules and excited states grouped into nine categories (A–I): (A) unsaturated hydrocarbons, (B) carbonyls, (C) sulfur analogues, (D) aromatic hydrocarbons and heterocycles, (E) amides and nitrogen-rich compounds, and (F) nucleobases, along with (G) donor–acceptor (D–A) and intramolecular short-range (SR) charge-transfer (CT) systems, (H) Rydberg states with diffuse electron density, and (I) singlet–triplet pairs of multireference character. Sets A–F cover the most optically relevant $\pi \to \pi^*$ and $n \to \pi^*$ excitations in 48 small- to medium-sized chromophores (Fig.~\ref{fig:molecules}), including 28 benchmark molecules from Thiel \textit{et al.}\cite{schreiber2008} and 20 systems from the QUEST database\cite{veril2021}. Sets G–I add 24 additional systems from the QUEST\cite{loos2018,veril2021} and Truhlar–Gagliardi benchmarks\cite{hoyer2015}, encompassing prototypical CT, Rydberg, and singlet–triplet excitations in aromatic, aliphatic, and heteroatom-containing systems. Altogether, 72 vertical excitations were evaluated and summarized in Tab~1 across methods relative to EOM-CCSD. Four singlet states—benzene, furan, naphthalene, and all-\textit{E}-hexatriene—are common to the categories A–F and I, resulting in 68 distinct electronic excitations.: 54 valence (28 $n \to \pi^*$, 26 $\pi \to \pi^*$), 5 CT, and 9 Rydberg states, spanning 59 singlets and 9 triplets (5 valence and 4 Rydberg excitations). This selection covers all major excitation classes of theoretical and practical importance—localized, diffuse, and strongly correlated. UGA-SSMRPT2 is equally capable of treating both singlet and triplet manifolds. The low-lying excited states are emphasized owing to their central importance in spectroscopy, photophysics, and optoelectronic function. Charge-transfer and Rydberg states present particular challenges for perturbative approaches, where intruder states involving diffuse orbitals can destabilize the expansion, unless special care is taken to formulate the working equations. The intruder avoidance without the use of level shifts is one of the main strength of Mukherjee's multireference state-specific methods\cite{MAHAPATRA1998AApplications} and the results reflect this robustness. 

\textcolor{black}{For multireference treatments of electronically excited states, an appropriate choice of the active space is essential and involves selecting a subset of molecular orbitals from an underlying single-reference SCF calculation. Since no unique or universally optimal prescription exists, active space selection necessarily relies on chemical and physical insight, and the development of systematic and automated protocols remains an active area of research.\cite{bao2019,golub2021,stein2019} Broadly, two complementary strategies are commonly employed: energy-based approaches, which select quasi-degenerate occupied and virtual orbitals, and phenomenon-based approaches, which target orbitals directly involved in a specific physical process, such as a given class of electronic excitations.\cite{king2021,veryazov2021} In practice, energy-based selection can become ambiguous for larger systems due to orbital reordering and the absence of clear energetic cutoffs, whereas phenomenon-based selection provides a more transparent and chemically motivated framework for excited-state calculations. In the present work, active spaces were selected using a combined energy-based and phenomenon-based protocol guided by natural orbital occupation numbers and orbital character analyses. For each targeted electronic state, all orbitals directly involved in the dominant excitation channels (e.g., $\pi \rightarrow \pi^*$) were included in a minimal chemically relevant active space, while orbitals of predominantly $\sigma/\sigma^*$ or diffuse Rydberg character were excluded unless significant mixing with the valence orbitals was observed. Candidate active spaces were assessed by explicitly monitoring the stability of CASSCF excitation energies upon systematic enlargement of the active space.}

\textcolor{black}{As a representative example, Table~S6 of the Supplementary Material reports excitation energies for the $1\,^1B_u$ and $2\,^1A_g$ states of {E}-butadiene obtained with progressively enlarged active spaces ranging from CAS(2,2) to CAS(4,8). The low-lying excited states of butadiene are well known to be dominated by valence $\pi \rightarrow \pi^*$ excitations. In particular, the HOMO$-1$, HOMO, and LUMO orbitals all possess clear $\pi/\pi^*$ character, leading naturally to a CAS(4,3) description. Extension beyond this minimal valence space was achieved by systematically identifying additional $\pi^*$ virtual orbitals that are energetically interleaved with $\sigma^*$ orbitals and explicitly swapping them into the active space based on orbital character analysis, thereby ensuring that all enlarged active spaces remain restricted to the valence $\pi/\pi^*$ manifold. Although relatively small active spaces such as CAS(4,4) may yield apparently reasonable CASSCF excitation energies, convergence of both the $1\,^1B_u$ and $2\,^1A_g$ states is only achieved once the full valence $\pi/\pi^*$ manifold is included, corresponding to active spaces in the range CAS(4,6)–CAS(4,8).}

\textcolor{black}{Larger active spaces employed in benchmark studies by Thiel et al\cite{silva-junior2010b} or Loos et al\cite{veril2021} incorporate additional $\sigma$ and diffuse orbitals to approach near-FCI accuracy; however, these orbitals primarily contribute dynamic correlation effects. As discussed later also, SSMRPT2 and NEVPT2 work best in terms of balance of static and dynamic correlation, when using a minimal active space unlike CASPT2. Moreover, given the steep scaling of UGA-SSMRPT2 with respect to active space size, the minimal chemically relevant valence manifold is the practical choice. We are convinced of its adequacy through the observed stabilization of excitation energies upon inclusion of dynamic correlation through the perturbative framework. The final active spaces employed for all systems considered in this work are summarized in Table~1. We note that although UGA-SSMRPT2 is not formally invariant with respect to active orbital rotations, it has been shown to be analytically size-consistent when localized molecular orbitals are employed, which is particularly important for the description of potential energy surfaces. In the present work, we use pseudo-canonical orbitals obtained from the state-averaged CASSCF calculations as implemented in GAMESS-US\cite{Schmidt1993}. The role of orbital choice, size consistency, and size extensivity has been discussed in detail in the context of potential energy curves in our previous works\cite{Sen2015,Sen2015b}.}

Vertical excitation energies for all categories (A–I) were computed and tabulated in Tab~1 using the newly developed UGA-SSMRPT2 method alongside strongly contracted NEVPT2 (SC-NEVPT2), fully internally contracted CASPT2 (FIC-CASPT2), and MCQDPT2. These established MRPT approaches are not considered as benchmarks but serve as important comparators that contextualize the accuracy and efficiency of UGA-SSMRPT2. In all cases, the same CASSCF reference wavefunctions were employed, \textcolor{black}{with active spaces chosen according to the protocol described above}. The chosen CAS spaces along with the corresponding CASSCF excitation energies are also reported in Tab~1. To ensure numerical stability and a balanced description of ground and excited states, state averaging over the two lowest singlets (S$_0$ and S$_1$) was applied, while the lowest triplet state (T$_1$) was treated without state averaging. The resulting excitation energies were directly compared against EOM-CCSD data in the same basis sets, which provides a reliable reference for systems dominated by single-excitation character and dynamic correlation. EOM-CCSD (although not always perfect) was chosen as the primary benchmark method to maintain consistency, as it enabled us to carry out all the benchmark computations in a controlled manner. Deviations from EOM-CCSD are discussed where relevant. Theoretical best estimates (TBE) from literature are also listed in Tab~1. They are discussed in case of ambiguity or unexpected disagreements among the theoretical methods considered. Notably, UGA-SSMRPT2 demonstrates robust performance even with compact active spaces, offering comparable accuracy to CASPT2 with larger active spaces and NEVPT2 with a two-body H$_0$.

\begin{center}
\begin{scriptsize}
\setlength{\tabcolsep}{0.75pt}  
\begin{longtable}{l l l c c c c c c c c}
\caption{Vertical excitation energies (in eV) for the chosen \textit{test set} comparing CASSCF, SSMRPT, $<$SSMRPT$>$, MCQDPT, SC-NEVPT2, FIC-CASPT2, EOM-CCSD and TBE. TBE values are taken from $^{a}$Ref\citenum{silva-junior2010b}, 
$^{b}$Ref\citenum{veril2021}, 
$^{c}$Ref\citenum{stein2009}, 
$^{d}$Ref\citenum{sun2013}, 
$^{e}$Ref\citenum{moore2013}, 
$^{f}$Ref\citenum{loos2018}, and
$^{g}$Ref\citenum{tomic2005}}. \\
\toprule
Molecule & State & CAS & CASSCF & SSMRPT & $\langle$SSMRPT$\rangle$ & MCQDPT & SC-NEVPT2 & FIC-CASPT2 & EOM-CCSD & TBE \\
\midrule
\endfirsthead
\toprule
Molecule & State & CAS & CASSCF & SSMRPT & $\langle$SSMRPT$\rangle$ & MCQDPT & SC-NEVPT2 & FIC-CASPT2 & EOM-CCSD & TBE \\
\midrule
\endhead
\midrule
\multicolumn{11}{r}{\textit{Continued on next page}} \\
\midrule
\endfoot
\bottomrule
\endlastfoot
\midrule
\multicolumn{11}{l}{\textbf{A. Unsaturated aliphatic hydrocarbons}} \\
Ethene & $^1$B$_{1u}$ (\(\pi \rightarrow \pi^*\)) & 2,2 & 9.04 & 8.47 & 8.46 & 7.91 & 8.12 & 7.77 & 8.33 & 7.80$^{a}$,7.93$^{b}$ \\
E-butadiene & $^1$B$_u$ (\(\pi \rightarrow \pi^*\)) & 4,4 & 7.24 & 6.43 & 6.46 & 6.22 & 6.26 & 5.87 & 6.56 & 6.18$^{a}$, 6.22$^{b}$ \\
all-E-hexatriene & $^1$B$_u$ (\(\pi \rightarrow \pi^*\)) & 6,5 & 6.28 & 5.31 & 5.35 & 5.11 & 5.20 & 4.84 & 5.57 & 5.10$^{a}$, 5.37$^{b}$ \\
all-E-octatetraene & $^1$B$_u$ (\(\pi \rightarrow \pi^*\)) & 2,4 & 5.47 & 4.65 & 4.66 & 4.31 & 4.49 & 4.07 & 4.92 & 4.66$^{a}$, 4.78$^{b}$ \\
Cyclopropene & $^1$B$_2$ (\(\pi \rightarrow \pi^*\)) & 4,3 & 7.24 & 6.86 & 6.80 & 6.54 & 6.63 & 6.36 & 6.87 & 6.68$^{a}$, 6.68$^{b}$ \\
Cyclopentadiene & $^1$B$_2$ (\(\pi \rightarrow \pi^*\)) & 4,4 & 6.28 & 5.60 & 5.64 & 5.22 & 5.42 & 5.04 & 5.71 & 5.55$^{a}$, 5.54$^{b}$ \\
Norbornadiene & $^1$A$_2$ (\(\pi \rightarrow \pi^*\)) & 4,4 & 5.78 & 5.61 & 5.63 & 5.18 & 5.40 & 4.97 & 5.66 & 5.37$^{a}$ \\
Acetylene & $^1\Sigma_u^-$ (\(\pi \rightarrow \pi^*\)) & 4,5 & 7.67 & 7.48 & 7.46 & 7.11 & 7.30 & 6.91 & 7.27 & 7.10$^{b}$ \\
Diacetylene & $^1\Sigma_u^-$ (\(\pi \rightarrow \pi^*\)) & 4,4 & 5.99 & 5.31 & 5.33 & 4.74 & 5.38 & 4.44 & 5.46 & 5.33$^{b}$ \\
Methylenecyclopropene & $^1$B$_2$ (\(\pi \rightarrow \pi^*\)) & 4,5 & 4.21 & 4.67 & 4.67 & 4.27 & 4.45 & 3.95 & 4.69 & 4.28$^{b}$ \\

\addlinespace
\multicolumn{11}{l}{\textbf{B. Aldehydes, ketones, and related carbonyls}} \\
Formaldehyde & $^1$A$_2$ (n $\rightarrow$ $\pi^*$) & 2,7 & 3.67 & 4.06 & 4.02 & 3.59 & 3.81 & 3.48 & 3.96 & 3.88$^{a}$, 3.98$^{b}$ \\
Acetaldehyde & $^1$A$''$ (n $\rightarrow$ $\pi^*$) & 6,4 & 4.20 & 4.58 & 4.52 & 4.29 & 4.48 & 4.09 & 4.39 & 4.31$^{a}$ \\
Acetone & $^1$A$_2$ (n $\rightarrow$ $\pi^*$) & 6,6 & 4.71 & 4.41 & 4.36 & 4.25 & 4.48 & 4.12 & 4.46 & 4.38$^{a}$, 4.47$^{b}$ \\
Cyclopropenone & $^1$B$_1$ (n $\rightarrow$ $\pi^*$) & 6,5 & 5.47 & 4.31 & 4.31 & 3.87 & 4.31 & 3.73 & 4.56 & 4.26$^{b}$ \\
Ketene & $^1$A$_2$ (\(\pi \rightarrow \pi^*\)) & 4,6 & 4.28 & 3.88 & 3.83 & 3.68 & 3.89 & 3.58 & 3.99 & 3.85$^{b}$ \\
Acrolein & $^1$A$''$ (n $\rightarrow$ $\pi^*$) & 8,8 & 4.10 & 3.87 & 3.82 & 3.62 & 3.85 & 3.50 & 3.92 & 3.78$^{b}$ \\
Glyoxal & $^1$A$_u$ (n $\rightarrow$ $\pi^*$) & 6,6 & 3.70 & 2.91 & 2.88 & 2.30 & 3.12 & 2.17 & 3.01 & 2.88$^{b}$ \\
Propynal & $^1$A$''$ (n $\rightarrow$ $\pi^*$) & 8,8 & 4.35 & 3.93 & 3.90 & 3.72 & 4.00 & 3.64 & 3.95 & 3.80$^{b}$ \\
\textit{p}-Benzoquinone & $^1$B$_{1g}$ (n $\rightarrow$ $\pi^*$) & 8,6 & 3.45 & 2.84 & 2.70 & 1.57 & 2.49 & 1.52 & 3.07 & 2.74$^{a}$, 2.82$^{b}$ \\

\addlinespace
\multicolumn{11}{l}{\textbf{C. Sulfur-containing molecules}} \\
Thioformaldehyde & $^1$A$_2$ (n $\rightarrow$ $\pi^*$) & 4,3 & 2.26 & 2.30 & 2.34 & 2.15 & 2.27 & 2.04 & 2.29 & 2.22$^{b}$ \\
Thioacetone & $^1$A$_2$ (n $\rightarrow$ $\pi^*$) & 6,6 & 3.19 & 2.55 & 2.59 & 2.42 & 2.61 & 2.37 & 2.62 & 2.53$^{b}$ \\
Thiopropynal & $^1$A$''$ (n $\rightarrow$ $\pi^*$) & 8,6 & 2.10 & 2.03 & 2.07 & 1.95 & 2.10 & 1.81 & 2.15 & 2.03$^{b}$ \\
Cyclopropenethione & $^1$A$_2$ (n $\rightarrow$ $\pi^*$) & 6,6 & 3.86 & 3.48 & 3.50 & 3.31 & 3.48 & 3.17 & 3.52 & 3.41$^{b}$ \\

\addlinespace
\multicolumn{11}{l}{\textbf{D. Aromatic hydrocarbons and heterocycles}} \\
Benzene & $^1$B$_{2u}$ (\(\pi \rightarrow \pi^*\)) & 6,6 & 7.03 & 5.18 & 5.20 & 4.29 & 4.84 & 3.98 & 5.20 & 5.06$^{a}$, 5.06$^{b}$ \\
Furan & $^1$B$_2$ (\(\pi \rightarrow \pi^*\)) & 4,3 & 7.25 & 6.61 & 6.54 & 6.04 & 6.54 & 6.11 & 6.66 & 6.32$^{a}$, 6.37$^{b}$ \\
Pyrrole & $^1$B$_2$ (\(\pi \rightarrow \pi^*\)) & 6,6 & 5.69 & 6.56 & 6.57 & 5.86 & 6.44 & 6.11 & 6.38 & 6.57$^{a}$, 6.26$^{b}$ \\
Pyridine & $^1$B$_1$ (n $\rightarrow$ $\pi^*$) & 8,6 & 5.06 & 5.05 & 5.08 & 4.82 & 4.69 & 3.94 & 5.19 & 4.59$^{a}$, 4.95$^{b}$ \\
Pyrazine & $^1$B$_{3u}$ (n $\rightarrow$ $\pi^*$) & 4,6 & 7.12 & 4.44 & 4.44 & 3.80 & 3.81 & 2.70 & 4.34 & 4.13$^{a}$, 4.15$^{b}$ \\
Pyrimidine & $^1$B$_1$ (n $\rightarrow$ $\pi^*$) & 8,8 & 5.10 & 4.51 & 4.46 & 4.07 & 4.55 & 3.99 & 4.65 & 4.43$^{a}$, 4.44$^{b}$ \\
Pyridazine & $^1$B$_1$ (n $\rightarrow$ $\pi^*$) & 8,8 & 5.05 & 3.89 & 3.93 & 3.65 & 4.09 & 3.54 & 4.05 & 3.85$^{a}$, 3.83$^{b}$ \\
s-Triazine & $^1$A$''_1$ (n $\rightarrow$ $\pi^*$) & 8,7 & 5.23 & 5.08 & 5.06 & 4.77 & 4.47 & 3.17 & 4.91 & 4.70$^{a}$, 4.72$^{b}$ \\
s-Tetrazine & $^1$B$_{3u}$ (n $\rightarrow$ $\pi^*$) & 6,8 & 7.18 & 2.60 & 2.64 & 1.10 & 1.90 & 1.02 & 2.66 & 2.46$^{a}$, 2.47$^{b}$ \\
Imidazole & $^1$A$'$ (\(\pi \rightarrow \pi^*\)) & 8,8 & 5.95 & 6.81 & 6.84 & 6.62 & 6.85 & 6.46 & 6.74 & 6.25$^{a}$, 6.41$^{b}$ \\
Thiophene & $^1$A$_1$ (\(\pi \rightarrow \pi^*\)) & 4,3 & 7.01 & 6.10 & 6.07 & 5.39 & 5.93 & 5.41 & 6.21 & 5.64$^{b}$ \\
Naphthalene & $^1$B$_{2u}$ (\(\pi \rightarrow \pi^*\)) & 4,3 & 6.10 & 4.46 & 4.43 & 4.03 & 4.26 & 3.82 & 4.40 & 4.82$^{a}$, 4.90$^{b}$ \\

\addlinespace
\multicolumn{11}{l}{\textbf{E. Amides and nitrogen-containing small molecules}} \\
Formamide & $^1$A$''$ (n $\rightarrow$ $\pi^*$) & 4,8 & 5.86 & 5.75 & 5.76 & 5.53 & 5.74 & 5.45 & 5.71 & 5.55$^{a}$, 5.65$^{b}$ \\
Acetamide & $^1$A$''$ (n $\rightarrow$ $\pi^*$) & 4,8 & 5.88 & 5.82 & 5.82 & 5.57 & 5.76 & 5.54 & 5.76 & 5.62$^{a}$ \\
Propanamide & $^1$A$''$ (n $\rightarrow$ $\pi^*$) & 4,8 & 5.91 & 5.84 & 5.85 & 5.62 & 5.82 & 5.57 & 5.78 & 5.65$^{a}$ \\
Nitrosomethane & $^1$A$''$ (n $\rightarrow$ $\pi^*$) & 8,7 & 2.28 & 1.96 & 1.95 & 1.76 & 1.92 & 1.54 & 1.98 & 1.96$^{b}$ \\
Diazomethane & $^1$A$_2$ (\(\pi \rightarrow \pi^*\)) & 6,6 & 4.00 & 3.16 & 3.14 & 2.88 & 3.13 & 2.80 & 3.21 & 3.14$^{b}$ \\
Cyanogen & $^1\Sigma_u^-$ (\(\pi \rightarrow \pi^*\)) & 8,8 & 6.29 & 6.35 & 6.39 & 6.47 & 6.48 & 5.75 & 6.53 & 6.39$^{b}$ \\
Streptocyanin C1 & $^1$B$_2$ (\(\pi \rightarrow \pi^*\)) & 4,3 & 8.02 & 7.32 & 7.37 & 6.76 & 7.33 & 6.79 & 7.44 & 7.13$^{b}$ \\
Cyanoacetylene & $^1\Sigma_u^-$ (\(\pi \rightarrow \pi^*\)) & 8,7 & 5.90 & 5.97 & 6.04 & 5.80 & 6.14 & 5.74 & 5.92 & 5.80$^{b}$ \\
Cyanoformaldehyde & $^1$A$''$ (n $\rightarrow$ $\pi^*$) & 8,8 & 4.38 & 4.05 & 4.02 & 3.75 & 3.98 & 3.64 & 3.95 & 3.81$^{b}$ \\

\addlinespace
\multicolumn{11}{l}{\textbf{F. Nucleobases}} \\
Adenine & $^1$A$'$ (\(\pi \rightarrow \pi^*\)) & 2,2 & 6.63 & 5.65 & 5.65 & 5.29 & 5.50 & 4.90 & 5.55 & 5.25$^{a}$ \\
Cytosine & $^1$A$'$ (\(\pi \rightarrow \pi^*\)) & 2,2 & 6.29 & 4.72 & 4.73 & 4.04 & 4.35 & 3.81 & 4.99 & 4.66$^{a}$,4.83$^{g}$ \\
Thymine & $^1$A$'$ (\(\pi \rightarrow \pi^*\)) & 2,2 & 6.79 & 5.27 & 5.28 & 4.73 & 4.98 & 4.42 & 5.16 & 5.20$^{a}$ \\
Uracil & $^1$A$''$ (n $\rightarrow$ $\pi^*$) & 2,2 & 7.01 & 5.61 & 5.60 & 4.86 & 5.17 & 4.65 & 5.14 & 5.00$^{a}$ \\

\addlinespace
\multicolumn{11}{l}{\textbf{G. Charge Transfer}} \\
BTCNE & $^1A_1 (\pi \to \pi^*)$ & 2,2 & 3.20 & 3.65 & 3.65 & 3.58 & 3.79 & 3.39 & 4.15 & 3.59$^{c}$ \\
DAN & $^1A_1 (\pi \to \pi^*)$ & 2,2 & 5.05 & 4.08 & 3.80 & 5.23 & 4.21 & 3.54 & 4.16 & 3.94$^{d}$ \\
DMABN & $^1A_1 (\pi \to \pi^*)$ & 2,2 & 6.09 & 4.75 & 4.74 & 3.88 & 4.42 & 3.85 & 4.65 & 4.72$^{e}$ \\
pNA & $^1A_1 (\pi \to \pi^*)$ & 2,2 & 5.79 & 4.46 & 4.45 & 3.92 & 4.44 & 3.73 & 4.15 & 4.39$^{f}$ \\
HCl & $^1\Pi$ (CT) & 6,5 & 7.62	& 7.97 & 7.96 & 7.77 & 7.79 & 7.67 & 7.91 & 7.84$^{f}$ \\
\addlinespace
\multicolumn{11}{l}{\textbf{H. Rydberg}} \\
Water & $^1B_1 (n \to 3s)$ & 6,8 & 7.31 & 7.69 & 7.70 & 7.58 & 7.57 & 7.53 & 7.60 & 7.62$^{f}$ \\
       & $^3B_1 (n \to 3s)$ & 6,8 & 6.78 & 7.38 & 7.39 & 7.30 & 7.20 & 7.15 & 7.20 & 7.21$^{f}$ \\
H$_2$S & $^1B_1 (n \to 4s)$ & 6,8 & 6.17 & 6.25 & 6.26 & 6.26 & 6.29 & 6.10 & 6.25 & 6.27$^{f}$ \\
       & $^3A_2 (n \to 4p)$ & 6,8 & 5.82 & 5.89 & 5.90 & 5.80 & 5.90 & 5.70 & 5.85 & 5.81$^{f}$ \\
NH$_3$ & $^1A_2 (n \to 3s)$ & 6,8 & 6.20 & 6.64 & 6.65 & 6.53 & 6.56 & 6.49 & 6.60 & 6.59$^{f}$ \\
       & $^3A_2 (n \to 3s)$ & 6,8 & 5.62 & 6.51 & 6.51 & 6.52 & 6.28 & 6.21 & 6.30 & 6.31$^{f}$ \\
Ethene & $^1B_{3u} (\pi \to 3s)$ & 2,2 & 6.13 & 7.47 & 7.47 & 7.36 & 7.45 & 7.32 & 7.31 & 7.45$^{f}$ \\
       & $^3B_{1u} (\pi \to \pi^*)$ & 2,2 & 5.09 & 4.64 & 4.64 & 4.56 & 7.33 & 7.21 & 4.46 & 4.55$^{f}$ \\
Isobutene & $^1A_1 (\pi, 3p_x)$ & 4,4 & 5.56 & 6.97 & 6.97 & 6.76 & 6.46 & 6.32 & 6.94 & 7.01$^{f}$ \\
\addlinespace
\multicolumn{11}{l}{\textbf{I. Singlet--Triplet pair}} \\
Benzene & $^1B_{2u} (\pi \to \pi^*)$ & 6,6 & 7.03 & 5.18 & 5.20 & 4.29 & 4.84 & 3.98 & 5.20 & 5.06$^{a}$ \\
        & $^3B_{2u} (\pi \to \pi^*)$ & 6,6 & 4.65 & 4.16 & 4.17 & 3.50 & 4.18 & 3.23 & 3.96 & 4.12$^{a}$ \\
Furan   & $^1B_2 (\pi \to \pi^*)$ & 4,5 & 7.19 & 6.70 & 6.68 & 6.04 & 6.48 & 5.95 & 6.66 & 6.32$^{a}$ \\
        & $^3B_2 (\pi \to \pi^*)$ & 4,5 & 3.65 & 4.39 & 4.40 & 4.12 & 4.37 & 3.89 & 4.08 & 4.17$^{a}$ \\
Naphthalene & $^1B_{3u} (\pi \to \pi^*)$ & 4,3 & 6.10 & 4.46 & 4.43 & 4.03 & 4.26 & 3.82 & 4.40 & 4.82$^{a}$ \\
           & $^3B_{3u} (\pi \to \pi^*)$ & 4,3 & 3.65 & 2.96 & 2.93 & 2.80 & 3.09 & 2.44 & 3.00 & 3.11$^{a}$ \\
All-E-Hexatriene & $^1B_u (\pi \to \pi^*)$ & 6,5 & 6.28 & 5.31 & 5.35 & 5.11 & 5.20 & 4.84 & 5.57 & 5.10$^{a}$ \\
           & $^3B_u (\pi \to \pi^*)$ & 6,5 & 3.07 & 2.54 & 2.57 & 2.18 & 3.14 & 2.75 & 2.64 & 2.69$^{a}$ \\
CO & $^1\Pi (n \to \pi^*)$ & 6,5 & 9.25 & 8.67 & 8.66 & 9.12 & 9.99 & 9.48 & 8.59 & 8.54$^{f}$ \\
   & $^3\Pi (n \to \pi^*)$ & 6,5 & 6.18 & 6.46 & 6.52 & 6.40 & 6.34 & 6.09 & 6.36 & 6.41$^{f}$ \\
\hline
\end{longtable}
\end{scriptsize}
\end{center}

\begin{figure}
    \centering
    \includegraphics[width=0.9\linewidth]{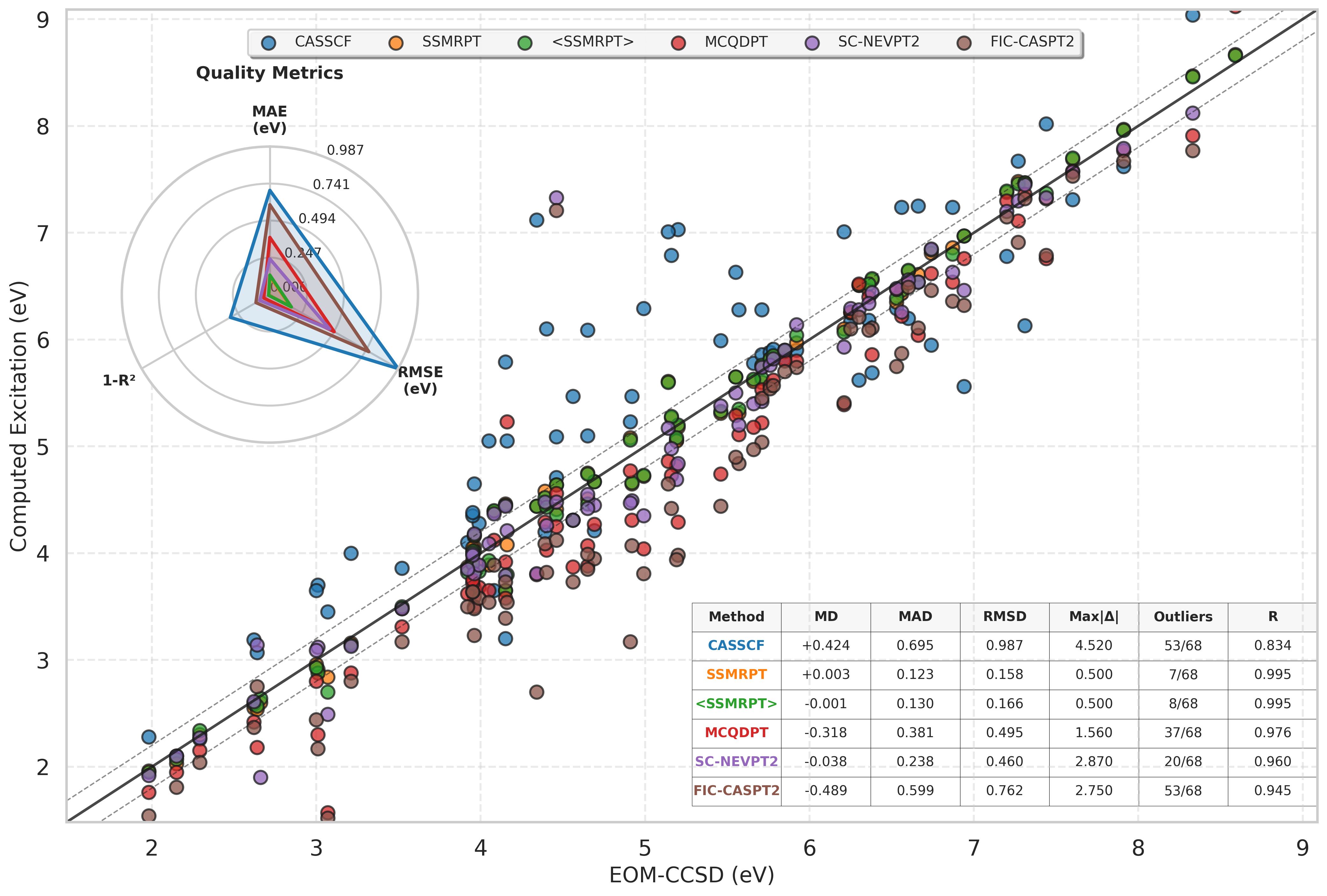}
    \caption{Correlation Plot for 68 excitations with all MRPTs vs EOM-CCSD with the error bar set at 0.25 eV}
    \label{fig:parity}
\end{figure}

\begin{figure}
    \centering
    \includegraphics[width=0.9\linewidth]{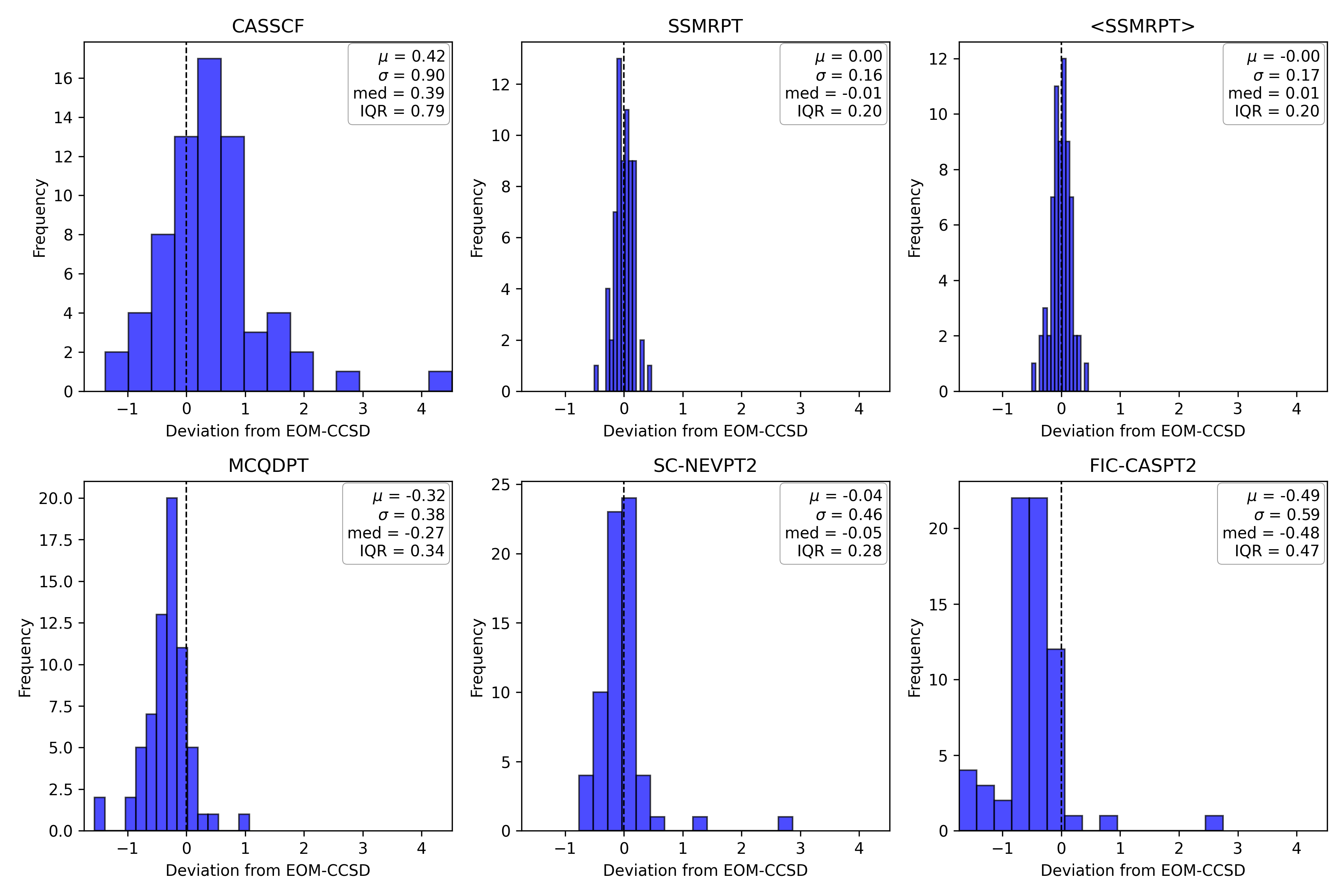}
    \caption{Signed Error histogram for 68 excitations with all MRPTs vs EOM-CCSD}
    \label{hist}
\end{figure}

\begin{figure}[H]
    \centering
    \includegraphics[width=\textwidth]{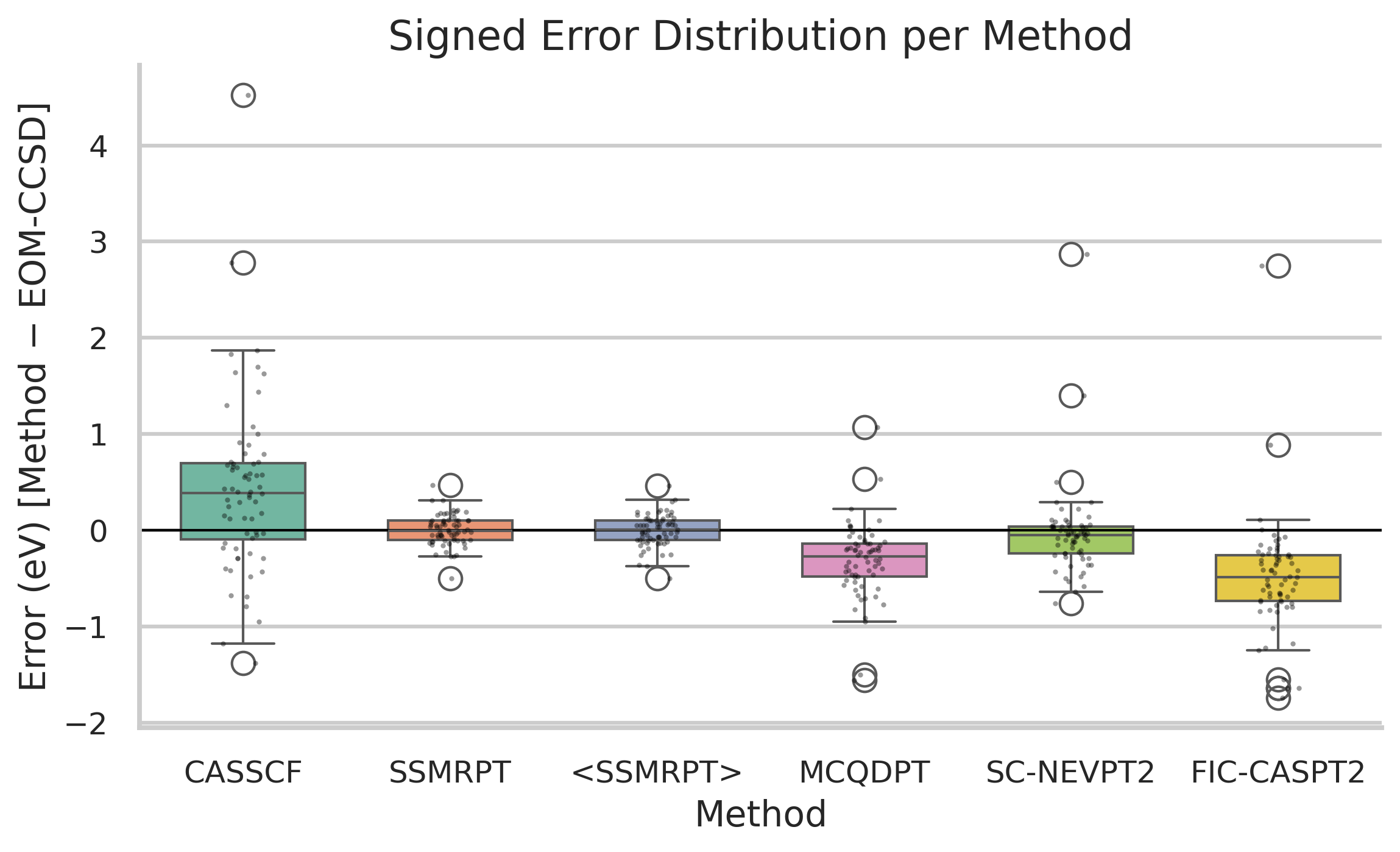}
    \caption{Box-and-whisker plots of signed errors ($E_{\mathrm{method}} - E_{\mathrm{EOM\text{-}CCSD}}$) for various wavefunction methods. Boxes represent the interquartile range (IQR) and medians, whiskers extend to 1.5×IQR, and circles denote outliers.}
    \label{fig:boxplot}
\end{figure}

A parity plot (Fig.~\ref{fig:parity}) comparing excitation energies computed with various methods against EOM-CCSD reveals clear performance hierarchies among multireference methods. UGA-SSMRPT2 and $<$UGA-SSMRPT2$>$ show near-perfect linearity ($R = 0.995$, $R^2 = 0.99$), negligible bias (MD = 0 eV), and minimal dispersion (MAD = 0.12–0.13 eV, RMSD = 0.16 eV), with nearly all points lying within $\pm$0.25 eV of the diagonal. MCQDPT and SC-NEVPT2 exhibit slight negative bias (MD = -0.32 and -0.04 eV) and broader scatter ($R$ = 0.98 and 0.96, $R^2$= 0.95 and 0.92, RMSD=0.50 and 0.46), while FIC-CASPT2 shows a stronger underestimation (MD = –0.49 eV) and an even wider scatter (RMSD = 0.76 eV, $R$=0.94 and $R^2$=0.89). CASSCF performs the poorest, with large positive deviations (MD = +0.42 eV), high variance (RMSD = 0.99 eV), and weak correlation ($R^2$ = 0.70). All the computed statistical quantities are tabulated in Tab.~2.
The corresponding signed-error histograms (Fig.~\ref{hist}) confirm these trends: SSMRPT-type methods yield sharp, symmetric, near-zero-centered distributions, corroborating their minimal standard deviation $\sigma$ ($\sim$0.16–0.17 eV). MCQDPT and NEVPT2 ($\sigma$ $\sim$0.38–0.46 eV) show moderately left-shifted profiles. 
FIC-CASPT2 exhibits a broader spread ($\sigma$ = 0.59 eV) with pronounced negative bias; while CASSCF displays the widest, right shifted distribution ($\sigma$ = 0.99 eV). Box-and-whisker statistics (Fig.~\ref{fig:boxplot}) further emphasize the superior robustness of SSMRPT approaches, with median errors near zero, narrow interquartile ranges (IQRs $\sim$0.20 eV), and only two moderate outliers (Uracil, 0.47 eV; BTCNE, 0.50 eV). In contrast, MCQDPT and NEVPT2 show noticeable negative bias (medians of -0.27 and -0.05 eV, respectively) and broader spreads (IQRs of 0.34 and 0.28 eV), each with four outliers. FIC-CASPT2 presents an even wider distribution (IQR = 0.48 eV), six outliers, and a systematic underestimation (median = -0.49 eV). CASSCF displays strong overestimation (median = +0.39 eV), the largest variance among all methods (IQR = 0.79 eV; maximum deviation = +4.5 eV), and three outliers. Collectively, these confirm that SSMRPT2 type methods achieve high precision and minimal systematic error with respect to EOM-CCSD reference data.

\begin{figure}
    \centering
    \includegraphics[width=1.0\linewidth]{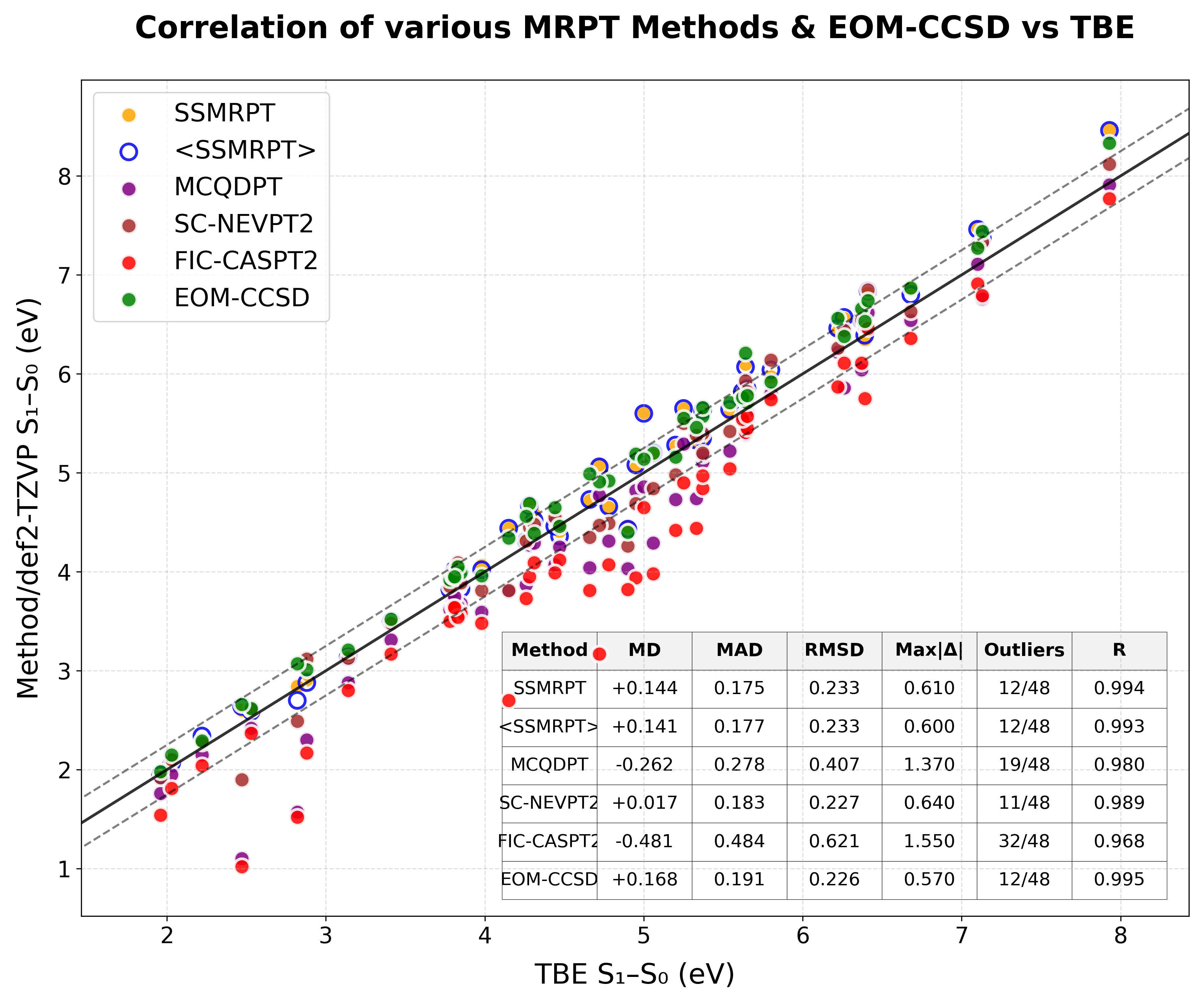}
    \caption{Correlation plot of S$_1$-S$_0$ (eV) for sets A--F using various methods with respect to TBEs from Loos et. al.\cite{veril2021} where available and from Thiel et al.\cite{silva-junior2010b} elsewhere. }
    \label{fig:parity_TBE}
\end{figure}

Theoretical Best Estimates (TBEs) from the literature were also examined. The TBE-1 dataset of Thiel et al.\cite{schreiber2008} (2008) for singlet states are predominantly derived from carefully optimized MS-CASPT2/TZVP performed with Molcas 6.4 or coupled cluster (CC) calculations with large basis sets, supplemented by MRCI, MRMP, and CC3/TZVP results where appropriate. In a subsequent study, Thiel et al.\cite{silva-junior2010b} (2010) refined their benchmark by reporting TBE-2 values for singlet states, defined primarily as CC3/aug-cc-pVTZ excitation energies except in cases where CC3/TZVP indicated less than 80 \% single-excitation character. For these strongly multireference cases, MS-CASPT2/aug-cc-pVTZ data were adopted instead. The authors noted that MS-CASPT2 exhibited occasional sensitivity to active-space definition and erratic behavior for some systems. In a few other cases, MRCI or MRMP energies were retained as TBE-2. Comprehensive discussions on methodological choices and dataset construction are available in Thiel et al.\cite{schreiber2008,silva-junior2010b}. The TBEs from the QUEST database (Loos and Jacquemin et al.)\cite{loos2018,loos2020,veril2021,loos2025}, are obtained from high level methods such as CCSDT, CCSDTQ, and even full configuration interaction (FCI) with large basis sets. These data serve as an independent cross-validation of the accuracy of both UGA-SSMRPT2 and EOM-CCSD especially in the cases of disagreement. Correlation plots comparing UGA-SSMRPT2, MC-QDPT, SC-NEVPT2, FIC-CASPT2, EOM-CCSD, and TBE—together with all raw numerical data are provided in the SI and a summary plot is shown in Fig.~\ref{fig:parity_TBE}.

\begin{table}[H]
\centering
\caption{Statistical analysis of the errors in excitation energy (in eV) relative to EOM-CCSD across all 68 excitations. MD: mean deviation, MAD: mean absolute deviation, RMSD: root-mean-square deviation, Max$|\Delta|$: maximum absolute deviation, IQR: interquartile range, $R$: Pearson correlation coefficient, $R^2$: coefficient of determination.}
\renewcommand{\arraystretch}{1.2}
\resizebox{1.05\textwidth}{!}{
\begin{tabular}{lrrrrrrrrrrrrr}
\hline
\textbf{Method} & MD & MAD & RMSD & Std & Median & Q1 & Q3 & IQR & Max$|\Delta|$ & Outliers & $R$ & $R^2$ \\
\hline
CASSCF            & +0.424 & 0.695 & 0.987 & 0.987 & 0.390 & -0.093 & 0.695 & 0.788 & 4.52 & 53 & 0.834 & 0.696 \\
\textbf{SSMRPT}   & +0.003 & 0.123 & 0.158 & 0.159 & -0.005 & -0.103 & 0.100 & 0.202 & 0.50 & 7  & 0.995 & 0.991 \\
$\langle$\textbf{SSMRPT}$\rangle$ & -0.001 & 0.130 & 0.166 & 0.167 & +0.005 & -0.100 & 0.100 & 0.200 & 0.50 & 8  & 0.995 & 0.990 \\
MCQDPT            & -0.318 & 0.381 & 0.495 & 0.382 & -0.270 & -0.483 & -0.140 & 0.342 & 1.56 & 37 & 0.976 & 0.953 \\
SC-NEVPT2         & -0.038 & 0.238 & 0.460 & 0.462 & -0.050 & -0.240 & 0.040 & 0.280 & 2.87 & 20 & 0.960 & 0.922 \\
FIC-CASPT2        & -0.489 & 0.599 & 0.762 & 0.589 & -0.485 & -0.733 & -0.257 & 0.475 & 2.75 & 53 & 0.945 & 0.893 \\
TBE-2             & -0.136 & 0.186 & 0.237 & 0.196 & -0.135 & -0.240 & -0.025 & 0.215 & 0.60 & 17 & 0.992 & 0.984 \\
\hline
\end{tabular}
}
\end{table}

We continue the discussions class by class, focusing on the states with errors larger than 0.25 eV wrt EOM-CCSD. A consideration of experimental and TBE values are made when the EOM-CCSD value deviates significantly from these values. A statistical survey of the 48 molecules, in sets A-F, demonstrates the excellent performance of UGA-SSMRPT2 for optical gaps. Unless specifically mentioned, numbers reported for UGA-SSMRPT2 refer to the relaxed variant. $<$UGA-SSMRPT2$>$ (unrelaxed) values typically lie within 0.05 eV of the relaxed numbers. The UGA-SSMRPT2 method for S$_1$-S$_0$ energies attains mean absolute deviation (MAD) relative to EOM-CCSD of 0.12 eV. The absolute errors, $|\Delta|$, lie within $\sim$0.20 eV with 7/48 outliers, within 0.25 eV with 4/48 outliers (Fig.~2) and 0.27 eV with 1/48 outlier. $<$UGA-SSMRPT2$>$ attains MAD relative to EOM-CCSD of 0.12 eV with absolute errors of 0.20 eV with 6/48 outlers, 0.25 eV with 4/48 outliers and 0.26 eV with 2/48 outliers. \textit{p}-benzoquinone shows a large difference of 0.14 eV between UGA-SSMRPT2 and $<$UGA-SSMRPT2$>$ values indicating a strong coupling between static and dynamic correlation. The unrelaxed value $<$UGA-SSMRPT2$>$ does not agree well with EOM-CCSD ($|\Delta| =$ 0.37 eV) while the relaxed value is within the error-bar ($|\Delta|$=0.23 eV). Excluding five persistent outliers (all-\textit{E}-hexatriene, octatetraene, cytosine, uracil, and p-benzoquinone), reduces the MAD to 0.09 eV, with no deviations exceeding 0.25 eV, and yields MD = 0.01 eV and RMSD = 0.11 eV over 43 molecules. 
This level of agreement is striking, particularly given that prior studies using TDDFT/B3LYP/cc-pVTZ reported typical MAD around 0.26 eV for the same dataset\cite{becke2018}. 
MADs for MCQDPT2, FIC-CASPT2 and SC-NEVPT2 wrt EOM-CCSD in sets A-F are 0.43, 0.65 and 0.19 eV respectively. The MAD of SC-NEVPT2 which uses a better H$_0$ is comparable to UGA-SSMRPT2. 

Uracil is a persistent outlier with the largest absolute deviation of $|\Delta|=$ 0.47 eV. For all-\textit{E}-octatetraene, all-\textit{E}-hexatriene, and cytosine, a large difference between UGA-SSMRPT2 and EOM-CCSD is not necessarily a deficiency of UGA-SSMRPT2 -- other possibilities include deficiencies in EOM-CCSD, basis set insufficiency due to diffuse or Rydberg character in the excited states, etc. Each case is discussed below. 

\textit{All-E-octatetraene:} The $^1$B$_u $ excited state of octatetraene shows an EOM-CCSD/TZVP excitation energy of 4.92 eV, in excellent agreement with the CC3/TZVP value of 4.94 eV\cite{schreiber2008}, confirming that the doubles excitation manifold adequately captures dynamic correlation. The corresponding CC3 \%T$_1$ value of 91.9\% places this transition firmly within the single-reference regime. Inclusion of diffuse and polarization functions significantly affects this state: CC3/6-31++G+2 yields 5.14 eV\cite{cronstrand2001}, about 0.2 eV higher than CC3/TZVP, whereas CC3/aug-cc-pVTZ provides a lower value of 4.84 eV\cite{silva-junior2010b}. The TBE-1 and TBE-2 values reported by Thiel et al. (4.66 eV) are based on the MRMP extrapolations of Nakayama et al.\cite{nakayama1998}, which involve a complex basis-set extrapolation procedure. The experimental 0–0 excitation energy is reported as 4.41 eV\cite{heimbrook1984}. The pronounced vibronic structure observed in the experimental spectrum indicates a significant contribution of the zero-point vibrational energy (ZPVE). Subsequent studies by Angeli and co-workers\cite{angeli2011} (2011), employing ANO basis sets and incorporating ZPVE corrections within the NEVPT2 framework, yielded vertical excitation energies of 4.7–4.9 eV depending on geometry. Adiabatic excitation energies at the same level of theory are then consistent with the experimental 0–0 value (4.41 eV)\cite{heimbrook1984}. They have, thus, identified the experimental value adopted by Thiel \text{et. al.}\cite{silva-junior2010b} as an adiabatic transition. CASPT2/CAS(8,16)+ZPVE computations give similar adiabatic values of 4.46–4.50 eV. Loos et al.\cite{veril2021} later reported a CCSDT/6-31+G(d) result of 4.78 eV (BSSE-corrected) as the most reliable vertical TBE for this state. Considering 4.78 eV as the TBE, UGA-SSMRPT2 (4.65 eV), and EOM-CCSD (4.92 eV) agree within 0.20 eV of the TBE.

\textit{All-E-hexatriene: } A nearly identical trend is observed for all-\textit{E}-hexatriene. The experimental vertical excitation energy for the $^1B_u$ state remains uncertain, and the value of 5.1 eV employed by Thiel et al. to identify the MRMP-extrapolated values of Nakayama et al.\cite{nakayama1998} as TBE-1 and TBE-2  
corresponds to the adiabatic 0–0 transition rather than the vertical excitation. More recent work by Guareschi and Angeli (2023)\cite{guareschi2023} has placed the true vertical excitation energy at 5.5–5.6 eV using NEVPT2 calculations, with the experimental 5.10 eV band\cite{flicker1977} reassigned as the adiabatic 0–0 energy after ZPVE correction. Loos et al.\cite{veril2021} further reported a high-level CCSDT/aug-cc-pVDZ value of 5.37 eV, providing a more reliable theoretical benchmark. Within this updated framework, UGA-SSMRPT2 (5.31 eV) and EOM-CCSD (5.57 eV) both lie within 0.20 eV of the new TBE.

\textit{Cytosine:} Gas phase experiments place the absorbtion maxima of cytosine at 4.5--4.65 eV\cite{tajti2009}. EOM-CCSD/TZVP (4.99 eV) severely overestimates the value presumably due to the strong multi-reference character of the excited states, considering that, EOM-CC3/cc-pVDZ\cite{tajti2009} gives a much lower value (different geometry from us) of 4.87 eV. 
MRPT/TZVP theories also give consistently lower values in the range 3.81--4.73 eV. DFT-MRCI/TZVP finds a value of 4.83 eV\cite{tomic2005}. The TBE-1 and TBE-2 of Thiel et al.\cite{schreiber2008,silva-junior2010b} is a CC2/aug-cc-pVTZ value of 4.66 eV. In view of the discussions above, the TBE can be safely considered in the range 4.6-4.8 eV wherein our UGA-SSMRPT2 value of 4.72 eV looks reasonable. The deviation in the EOM-CCSD value appears to be an inability to capture the multi-reference nature of the excited states. Furthermore, the HOMO-LUMO excitation only constitutes 78\% of the CASSCF states when a (2,2) active space is used. 

\textit{Uracil:} The TBE-2 reported by Thiel \textit{et al.}\cite{silva-junior2010b} for the $^1$A'' state of uracil is 5.00 eV computed with CREOM-CCSD(T)/aug-cc-pVTZ. CC2/aug-cc-pVTZ and CC2/aug-cc-pVQZ values are 4.81 eV and 4.80 eV respectively. CASPT2/TZVP with a (2,2) CAS gives 4.65 eV and with a (10,8) CAS gives 4.91 eV. The dominant single excitation contributes about 70\% of the CASSCF function in a (10,8) CAS indicating a very large multireference character\cite{silva-junior2010b}. It appears that perhaps the minimal (2,2) CAS adopted for our computations is insufficient for capturing the static correlation for this molecule. Our pilot code and computational resources were not sufficient for carrying out UGA-SSMRPT2 computations for uracil with an active space larger than (2,2).


An evaluation of the relative performance of the MRPT methods and EOM-CCSD for the various states considered, can be interpreted in terms of (a) magnitude of the correlation energy and its partitioning between static and dynamic, (b) multi-reference character of the states involved, (c) suitability of the active space chosen, and (d) suitability of the H$_0$ and the excitation space used for modelling the dynamic correlation. As expected, CASSCF (and also other mean field methods) tends to overestimate excitation energies, while correlated methods generally lower them. This trend holds for most systems in the benchmark set, with a few exceptions such as methylenecyclopropene, formaldehyde, acetaldehyde, pyrazole, imidazole, and cyanogen where excitation energies become higher on inclusion of dynamic correlation. Several molecules—cyclopropenone, all-\textit{E}-hexatriene, benzene, pyrazine, pyridazine, s-tetrazine, naphthalene, adenine, cytosine, thymine, and uracil—exhibit large differential correlation effects ($>$1 eV), indicating the need to capture differential correlation properly. \textcolor{black}{These pronounced effects can be traced back to qualitative differences in the spin or spatial symmetry and electronic character of the ground and excited states. For example, in all-\textit{E}-hexatriene the ground state is predominantly neutral, whereas the lowest optically allowed HOMO $\rightarrow$ LUMO excited state has significant ionic character. Similar changes in the balance between neutral, ionic, and charge-transfer configurations occur in several of the molecules listed above, leading to markedly different correlation energies between the two states.} In these cases, most MRPTs deviate by $>$1 eV relative to EOM-CCSD, whereas UGA-SSMRPT2 shows a maximum deviation of 0.47 eV for uracil (due to insufficient CAS size as discussed earlier) and remains within 0.27 eV for the remaining 47 molecules. SC-NEVPT2 performs poorest for cytosine ($\Delta = -0.94$ eV), MCQDPT2 for s-tetrazine ($\Delta = 1.55$ eV) and FIC-CASPT2 for s-triazine ($\Delta = 1.74$ eV). MCQDPT2 and FIC-CASPT2 exhibit errors above 1 eV for 2 and 7 molecules, respectively. 

\textcolor{black}{FIC--CASPT2\cite{anderson1990,forsberg1997,roos1995} results may be significantly improved by artificially inflating the active space in order to reduce the burden on the perturbative treatment. Indeed, FIC--CASPT2 with large active spaces can sometimes outperform even EOM--CCSD in approaching experimental excitation energies, especially when combined with empirical adjustments such as the IPEA shift\cite{forsberg1997}. In contrast, MCQDPT2\cite{hirao1992} exhibits severe convergence difficulties when the active space is artificially enlarged, as does UGA--SSMRPT2\cite{Sen2015,Sen2015b}. SC--NEVPT2\cite{angeli2001}, while formally convergent, often shows erratic behavior with excessively large active spaces, particularly when these exceed what is dictated by near-degeneracy considerations and physical intuition. Consequently, proponents of NEVPT2 and UGA--SSMRPT2 (MkMRPT2) have traditionally advocated the use of minimal, physically motivated active spaces, whereas FIC--CASPT2 studies frequently employ substantially larger spaces for the same molecular systems.
A representative example is provided by the $n \rightarrow \pi^*$ excitation to the $^1A_1$ state of \emph{s}-triazine. Thiel \emph{et al.}\cite{schreiber2008} recommended a CAS(12,9), whereas the minimal active space based on orbital near degeneracy is CAS(8,7). Using MS--CASPT2 with CAS(12,9), Thiel \emph{et al.} obtained an excitation energy of 4.66~eV, while FIC--CASPT2 with CAS(8,7) yielded only 3.17~eV. In comparison, UGA--SSMRPT2 with CAS(8,7) gave 5.08~eV and EOM--CCSD yielded 4.91~eV, against a CC3/TZVP reference value of 4.70~eV reported as TBE--2\cite{silva-junior2010b}. These results illustrate that FIC--CASPT2 generally requires substantially larger active spaces to achieve excitation energies of comparable accuracy.}


\textcolor{black}{It is important to note that the CASPT2 excitation energies reported in this work do not include the IPEA shift. Although empirical, the IPEA shift is now well established as a crucial correction for improving the accuracy of CASPT2 vertical excitation energies. In a comprehensive benchmark study of organic molecules using an aug-cc-pVTZ basis, Sarkar \emph{et al.}\cite{sarkar2022} demonstrated that CASPT2 strongly benefits from inclusion of the standard IPEA shift of 0.25~a.u. For a safe subset of 265 excited states, CASPT2 with IPEA exhibited a small positive MD of +0.06~eV and MAD of 0.11~eV, whereas CASPT2 without IPEA showed a pronounced systematic underestimation of excitation energies (MD of $-0.26$~eV) together with substantially larger scatter (MAD of 0.27~eV). This improvement was attributed to the IPEA shift correcting an imbalance in the CASPT2 zeroth-order Hamiltonian between open- and closed-shell configurations, which otherwise leads to overcorrelation of excited states and artificially lowered excitation energies. With IPEA, CASPT2 typically yields a modest blueshift relative to theoretical best estimates, but with significantly reduced dispersion.
In terms of overall performance, Sarkar \emph{et al.} found CASPT2 with IPEA to be comparable in accuracy to partially contracted NEVPT2 and competitive with widely used single-reference approaches such as ADC(2), CC2, and CCSD for states dominated by single-excitation character, while remaining inferior to high-level CC3. CASPT2 without IPEA was consistently inferior overall, except in certain Rydberg excitations where omission of the shift could occasionally be beneficial. The study further emphasized the basis-set dependence of the IPEA correction, noting that its benefits are most pronounced for triple- and quadruple-$\zeta$ basis sets, whereas unshifted CASPT2 may perform more favorably with smaller double-$\zeta$ bases. While no universally optimal IPEA value exists, 0.25~a.u.\ emerges as a robust and practical default. Since the primary focus of the present work is the assessment of UGA--SSMRPT2, and the effects of the IPEA shift on CASPT2 have already been extensively quantified in the literature, we restrict ourselves here to unshifted CASPT2 results and refer the reader to Ref.~\cite{sarkar2022} for a detailed analysis.}

However, sometimes large active spaces are physically necessary. The $^1B_{2u}$ excited state of naphthalene presents an instructive case. Thiel et al.\cite{schreiber2008,silva-junior2010b} reported TBE-1 = 4.77 eV (MS-CASPT2/TZVP, CAS(10,10)) and TBE-2 = 4.82 eV (CC3/TZVP with CCSDR(3) basis corrections), while Loos et al.\cite{veril2021} provided a TBE of 4.90 eV (CCSDT/6-31+G(d), BSSE-corrected). Our EOM-CCSD/TZVP value of 4.40 eV underestimates these benchmarks, as do the MRPT results (3.82–4.46 eV), with UGA-SSMRPT2 giving 4.46 eV. This consistent underestimation across single- and multireference formalisms likely stems from the inadequate capture of static correlation in EOM-CCSD and through the minimal CAS (4,3) used in the present MRPT computations. 
Enlarging the active space would be expected to recover the missing correlation and align the MRPT results with the established TBE of 4.8–4.9 eV. \textcolor{black}{
Another noteworthy case is the singlet $2\,^1A_g$ excited state of \emph{E}-butadiene, which is known to present a significant challenge for single-reference methods. For this state, EOM--CCSD substantially overestimates the excitation energy (7.45~eV), whereas higher-level EOM--CCSD(T) calculations from Ref\cite{watts1996a} report values in the range of 6.76--6.92~eV and CC3 values reported in Ref\cite{lehtonen2009} is 6.58 eV. As shown in Table~S6, the SSMRPT and $\langle$SSMRPT$\rangle$ excitation energies of 6.83 and 6.92~eV with a (4,4) CAS, respectively, are in close agreement with the EOM--CCSD(T) results\cite{watts1996a}, while CASSCF and MCQDPT underestimate the excitation energy. This observation highlights the limitations of relying on lower-level reference methods for strongly correlated excited states and motivates a reassessment of excitation energies with respect to reliable theoretical best estimates (TBE), as reflected in the statistical comparison against TBE values (Table~S4) presented in the Supplementary Material.
}

Using an updated set of TBEs—Loos \textit{et al.}~\cite{veril2021} where available and Thiel \textit{et al.}~\cite{silva-junior2010b} otherwise—we obtain an additional benchmark of all perturbative MR methods and EOM-CCSD for 48 singlet excitations (Sets A-F) presented in Fig.~\ref{fig:parity_TBE}. SSMRPT shows excellent agreement with the TBEs, with MD=+0.14 eV, MAD=0.18 eV, RMSD=0.23 eV, and 12 moderate outliers. The correlation coefficients (R=0.994, R$^2$=0.987) match or exceed those of SC-NEVPT2 (R=0.989, R$^2$=0.978). EOM-CCSD remains the most correlated method overall (R = 0.995, R$^2$=0.990), though its MD = +0.17 eV indicates a slightly stronger positive bias. MCQDPT shows a systematic negative shift (MD = –0.262 eV) and a significantly larger RMSD = 0.407 eV with 19 outliers, while FIC-CASPT2 exhibits the weakest performance using the same CAS, with large negative bias and dispersion (MD=-0.48 eV, MAD = 0.49 eV, RMSD = 0.63 eV) and 32 outliers. Overall, these statistics confirm that SSMRPT2 achieves accuracy and robustness comparable to the best-performing benchmarks, with minimal systematic bias and consistently high correlation with TBE values. The larger statistical deviations and increased outliers of UGA-SSMRPT2/TZVP can be traced to deficiencies in the basis set, as the TBE values refer to basis set extrapolated values in several cases.

Charge-transfer (CT) excitations (set G) present a stringent test for any perturbative formalism because they probe the coupling between localized donor–acceptor orbitals and long-range electron–hole separation, conditions under which most MRPT variants exhibit strong denominator instabilities. The present subset includes B-TCNE, DAN, DMABN, p-nitroaniline (pNA), and the prototypical HCl $\to$ H$^+$Cl$^-$ transition, covering both intramolecular short-range and intermolecular long-range CT regimes. Across the series, UGA-SSMRPT2 reproduces EOM-CCSD excitation energies with a MAD of 0.21 eV and a small MD of –0.02 eV, indicating a slight systematic underestimation but overall quantitative agreement (R = 0.985, R$^2$ = 0.97, RMSD = 0.27 eV). For the HCl CT transition—the canonical benchmark for long-range charge separation—UGA-SSMRPT2 predicts 7.97 eV, in near-perfect agreement with EOM-CCSD (7.91 eV) and TBE (7.84 eV)\cite{loos2018}, while CASPT2 and NEVPT2 deviate by 0.2–0.3 eV owing to residual intruder mixing and size-consistency loss. BTCNE and pNA are outliers with errors, $\Delta > 0.25$ eV against EOM-CCSD. 

\textit{BTCNE and pNA:} The UGA-SSMRPT2 excitation energies (3.65 eV for BTCNE and 4.46 eV for pNA) closely match the corresponding TBE\cite{stein2009,sun2013}(3.59 eV and 4.39 eV, respectively) but slightly different from the EOM-CCSD results (4.15 eV and 4.15 eV), leading to a maximum deviation of $\sim$0.5 eV. This suggests that the residual discrepancy primarily arises from the incomplete description of static correlation in EOM-CCSD rather than limitations of the perturbative model. Recently, Kozma \textit{et al.}\cite{kozma2020} reported that EOM-CCSD systematically overestimates CT excitation energies, typically by 0.2–0.3 eV relative to CCSDT-3, while STEOM-CCSD shows a clear improvement (MD $\sim$ 0 eV, RMSD $\sim$ 0.1 eV, and $R^2 > 0.98$). For BTCNE, the experimental CT energy of 3.59 eV\cite{hanazaki1972} (also solvent-influenced) is consistent with TDDFT/cc-pVDZ results obtained using optimally tuned range-separated functional (BNL) (3.8 eV)\cite{stein2009} and with SOS-ADC(2)/cc-pVTZ values from Kállay \textit{et al.}\cite{mester2022} (3.51 eV). Considering the range 3.5--3.8 eV, the UGA-SSMRPT2 (CAS(2,2)) prediction of 3.65 eV aligns well with both theoretical and experimental benchmarks. For \textit{p}NA, SA-CASSCF(12,12)/CASPT2 yields an excitation energy of 4.62 eV\cite{hoyer2015}, while $\delta$-CR-EOMCC(2,3)D\cite{kowalski2004}, calculations in 6-31G** basis set by Hoyer \textit{et al.}\cite{hoyer2015} report 4.30 eV. RI-CC2\cite{} results from Autschbach \textit{et al.}\cite{sun2013} give 4.39 eV, which is adopted here as TBE and the experimental charge-transfer (CT) band appears at 4.28 eV\cite{willetts1992}. Considering solvent effects, the best estimate for the TBE lies between 4.3–4.5 eV, a range accurately reproduced by both UGA-SSMRPT2 and SC-NEVPT2 using a minimal (2,2) CAS. 
Overall, it can be concluded that UGA-SSMRPT2 captures both short- and long-range charge-transfer excitations with uniform reliability.

Rydberg excitations (set H), involving H$_2$O, H$_2$S, NH$_3$, ethene, and isobutene, test the treatment of diffuse correlation and orbital relaxation in multireference perturbation theories owing to the highly diffuse character of the $n \to ns/np$ and $\pi \to ns/np$ transitions. Traditional MRPT2 approaches often struggle in this regime—compact active spaces typically lead to over- or under-correlation, and level shifts are commonly invoked to suppress intruder states arising from diffuse orbitals. In contrast, UGA-SSMRPT2 achieves quantitative agreement with EOM-CCSD for all members of the Rydberg subset using modest active spaces that include the essential lone-pair and diffuse orbitals. The MAD relative to EOM-CCSD is only 0.10 eV (RMSE = 0.13 eV), with a small positive mean deviation of +0.10 eV, indicating a uniform and slight overestimation but an excellent linear correlation (R$^2$=0.98) across the series. For instance, the $^1B_1(n \to 3s)$ transition in H$_2$O is predicted at 7.69 eV versus 7.60 eV (EOM-CCSD) and 7.62 eV (TBE)\cite{loos2018}, while the $^1A_2(n \to 3s)$ excitation in NH$_3$ is reproduced within 0.04 eV of both EOM-CCSD and TBE values. Even for the more diffuse $\pi \to 3p_x$ excitation in isobutene, UGA-SSMRPT2 (6.97 eV) remains within 0.03 eV of EOM-CCSD (6.94 eV), a level of precision unmatched by CASPT2 or NEVPT2 under similar active-space constraints. 

Accurate descriptions of singlet–triplet excitation pairs (set I) constitute one of the most stringent benchmarks for multireference perturbation theories, as these states require an accurate balance between static and dynamic correlation within near-degenerate $\pi$–manifolds. Set~I spans both aromatic systems (benzene, furan, naphthalene) and conjugated chains (all-\textit{E}-hexatriene) along with the heterodiatomic CO, covering diverse bonding topologies and correlation patterns. In this set, UGA-SSMRPT2 reproduces EOM-CCSD reference energies with a MAD of only 0.14 eV and MD of +0.09 eV (RMSE = 0.17 eV, R$^2$=0.994). For benzene, the canonical test for balanced $\pi \to \pi^*$ correlation, UGA-SSMRPT2 yields 5.20 eV for the $^1B_{2u}$ state and 4.17 eV for its triplet counterpart, in near-perfect accord with EOM-CCSD (5.20 / 3.96 eV) and TBE\cite{silva-junior2010b} values (5.06 / 4.12 eV). Similar sub-0.1 eV deviations are obtained for furan and naphthalene, demonstrating that the method maintains consistent singlet–triplet gaps without over-stabilizing either manifold—a common failure of CASPT2, NEVPT2, and MCQDPT when minimal active spaces are used. In view of the fact that the excitation energies of the $^1B_{2u}$ and $^3B_{2u}$ states of napthalene are not quantitatively reproduced by either UGA-SSMRPT2 (CAS(4,3)) or EOM-CCSD, the accurate recovery of the singlet–triplet gap (STG) by UGA-SSMRPT2 highlights its excellent error cancellation, arising from a balanced and state-specific treatment of static and dynamic correlation.
In extended systems such as all-\textit{E}-hexatriene, where correlation and near-degeneracy effects become pronounced, UGA-SSMRPT2 preserves the singlet–triplet separation within 0.1–0.2 eV of EOM-CCSD, correctly reproducing the experimental trend of decreasing gap with conjugation length. Even for CO—one of the most challenging $n \to \pi^*$ transitions with strong static correlation—the predicted 8.66 eV ($^1\Pi$) and 6.52 eV ($^3\Pi$) values match EOM-CCSD (8.59 / 6.36 eV) to within 0.1 eV, outperforming other MRPT variants that tend to overestimate this splitting by several tenths of an eV. Overall, UGA-SSMRPT2 provides a uniformly accurate and intruder-free treatment of both valence and mixed $n \to \pi^*$ singlet–triplet excitations. 

In summary, UGA-SSMRPT2 achieves a MAD of 0.12 eV and an RMSD of 0.16 eV relative to EOM-CCSD (R$^2$ = 0.99) across sets A--I. Category-wise analysis reveals consistently small errors (MAD = 0.06–0.14 eV) for valence, Rydberg, and singlet–triplet subsets (A–E, H, I), with near-perfect linear correlations (R $>$ 0.996). Sets F (MAD=0.24 eV) and G (MAD=0.21 eV) emerge as the most demanding: the nucleobase series (F) exhibits the largest deviation, driven by the strong multireference nature of $\pi \to \pi^*$ and $n \to \pi^*$ excitations in cytosine and uracil, where the compact (2,2) CAS used by us is actually smaller than the minimal CAS space required to capture the multi-configurational character. The numbers for all-\textit{E}-hexatetraene, all-\textit{E}-octatetraene and napthalene are also expected to improve with larger CAS choices. 
This is not a limitation of the theory per se but rather of our pilot code\cite{chakravarti2021}, which finds it difficult to handle the $n_{CAS}$ scaling with active space size (the same challenge exists for CASPT2/NEVPT2, though more technically advanced codes are available). Nonetheless, by avoiding empirical level shifts and intruder 
instabilities, UGA-SSMRPT2 offers a stable, computationally economical alternative to CASPT2/NEVPT2/MCQDPT for a wide range of excited-state problems where minimal active spaces provide balanced static/dynamic correlation treatment.
This balance of precision, efficiency, and generality positions UGA-SSMRPT2 as a powerful and scalable tool for excited-state studies in organic photophysics and optoelectronics and justifies further efforts to improve the computational capacity of the code.


\section{Summary and Future Outlook}

In addition to the earlier success of UGA-SSMRPT2 for computing potential energy surfaces\cite{Sen2015b}, the present benchmark study establishes UGA-SSMRPT2\cite{Sen2015,Sen2015b,chakravarti2021} as a quantitatively reliable and computationally efficient framework for excited-state electronic-structure calculations. Across 68 representative excitations encompassing $\pi \to \pi^*$, $n \to \pi^*$, charge-transfer, and Rydberg states, the method reproduces EOM-CCSD reference energies with overall MAD = 0.12 eV, RMSD = 0.16 eV, and $R^2$ = 0.99, signifying near-chemical accuracy using compact active spaces. In some cases where EOM-CCSD is insufficient, careful consideration of other methods have revealed that UGA-SSMRPT2 can capture better physics, thereby reproducing TBE values. Its state-specific multireference formulation intrinsically eliminates intruder-state problems and avoids empirical parameters such as the IPEA shifts of CASPT2, ensuring consistent performance across diverse excitation types.

UGA-SSMRPT2 achieves MAD = 0.06--0.14 eV ($R^2$ = 0.98--0.99) for all valence and Rydberg excitations, and MAD $<$ 0.14 eV for nearly all molecular groups (A–E, H, I) with excellent linear correlation ($R >= 0.99$). Charge-transfer (G) excitations remain the most demanding, yet UGA-SSMRPT2 attains MAD = 0.21 eV with minimal systematic bias (MD = -0.02 eV), demonstrating robustness even for long-range electron–hole separation. Deviations in nucleobases (F) arise from the strong multiconfigurational character of these states; nonetheless, all errors remain below 0.27 eV (except uracil) without empirical correction, underscoring the method’s numerical stability and generality. Collectively, these results confirm UGA-SSMRPT2’s ability to describe both localized and delocalized excitations within a unified perturbative framework, providing encouragement for further development.

At present, the formalism employs a class-diagonal one-body Fock-like $H_0$, analogous to Møller–Plesset\cite{moller1934} partitioning, though a diagonal variant exists\cite{Mao2012,Mao2012b}. Future improvements may incorporate the Epstein–Nesbet (EN)\cite{nesbet1955} partitioning or a two-body Dyall $H_0$\cite{dyall1995}, thereby recovering additional reference-space correlation while maintaining compact active spaces, akin to NEVPT2\cite{angeli2001}. Furthermore, implementation of the full exponential $e^{T_1}$ cluster operator—rather than its linearized form—could systematically enhance the accuracy of the perturbative expansion with regard to orbital relaxation. These represent natural extensions of the present formulation and constitute ongoing work. 

In conclusion, with technical improvements to handle large active spaces within computational limitations, UGA-SSMRPT2 shows promise in becoming a scalable and accurate method for computation of electronic excited states and potential energy surfaces alike.
 

\section*{Supporting Information}
Supporting information (SI) consists of the definitions of various statistical parameters discussed in this study, data corresponding to the plots in the manuscript, along with absolute error heatmap for all MRPTs wrt EOM-CCSD and additional statistical analyses for all the MRPT methods and EOM-CCSD wrt TBEs.

\section*{Acknowledgements}
We dedicate this article to Prof. Debashis Mukherjee on the occasion of his 79th birthday, in recognition of his pioneering contributions to many-body theory and quantum chemistry and his unwavering encouragements. S.S. gratefully acknowledges IISER Kolkata for infrastructure. S.C. and P.B. thank IISER Kolkata and the University Grants Commission (UGC), respectively, for their research fellowships. We also acknowledge the National Supercomputing Mission (NSM) for providing computational resources at PARAM RUDRA, hosted at the S. N. Bose National Centre for Basic Sciences, implemented by C-DAC, and supported by the Ministry of Electronics and Information Technology (MeitY) and the Department of Science and Technology (DST), Government of India.


\end{document}